\documentclass[aps,pra,superscriptaddress,twocolumn]{revtex4-2}
\usepackage{amssymb,amsmath,bbm,ragged2e,graphicx,physics,enumerate,subcaption,float,caption}
\usepackage[export]{adjustbox}

\begin{document}
\title{Maximum power of coupled-qubit Otto engines}
\author{Jingyi Gao}
\affiliation{Department of Physics, the University of Tokyo\\5-1-5 Kashiwanoha, Kashiwa, 277-8574, Japan}
\author{Naomichi Hatano}
\affiliation{Institute of Industrial Science, the University of Tokyo\\5-1-5 Kashiwanoha, Kashiwa, 277-8574, Japan}
\date{\today}
	
\begin{abstract}
	We put forward four schemes of coupled-qubit quantum Otto machine, a generalization of the single-qubit quantum Otto machine, based on work and heat transfer between an internal system consisting of a coupled pair of qubits and an external environment consisting of two heat baths and two work storages. The four schemes of our model are defined by the positions of attaching the heat baths, which play a key role in the power of the coupled-qubit engine. Firstly, for the single-qubit heat engine, we find a maximum-power relation, and the fact that its efficiency at the maximum power is equal to the Otto efficiency, which is greater than the Curzon-Ahlborn efficiency. Second, we compare the coupled-qubit engines to the single-qubit one from the point of view of achieving the maximum power based on the same energy-level change for work production, and find that the coupling between the two qubits can lead to greater powers but the system efficiency at the maximum power is lower than the single-qubit system's efficiency and the Curzon-Ahlborn efficiency.
\end{abstract}
\maketitle

\section{Introduction}
Quantum thermal machines \cite{PhysRevLett.127.200602, PhysRevResearch.4.033103, PhysRevA.102.062422} are attracting much attention recently, not only for their better performance than classical machines, but also for their value on exploring the potential applications of several quantum theories in different fields, such as quantum information and quantum thermodynamics. In particular, the quantum heat engine \cite{PhysRevResearch.3.023078, PhysRevA.106.022436, PhysRevA.102.042217, PhysRevE.105.034101, PhysRevLett.128.090602, PhysRevLett.128.180602} occupies an important position for its broad application scenarios and development prospects. 

Preliminary analyses of characteristics of quantum engines have been made in previous researches, especially for work production and efficiency \cite{PhysRevE.103.062109, PhysRevA.103.062225, PhysRevD.104.L041701, PhysRevB.104.125445}. Some quantum heat engines have been put forward in these years under the assumption of Maxwell's demon, validating a series of quantum information theories and their applications on the quantum heat engine \cite{PhysRevLett.123.250606, PhysRevLett.93.140403, PhysRevLett.124.100603}. Another aspect of quantum heat engine is given by quantum thermodynamics. Theory of open quantum systems \cite{NYSpringerVerlag, LondonAcademicPress, OxfordUniversityPress} plays a key role for quantum thermal machines by quantifying the evolution and simulating the interaction between the internal system and the external environment.

Improving the efficiency of quantum heat engines and the coefficient of performance for quantum refrigerators based on the Otto cycle or the Carnot cycle are the most attractive topics in these years \cite{PhysRevResearch.3.023078, PhysRevE.90.022102, PhysRevE.102.062123, CarlMBender_2000}. In contrast, the power has been less studied  \cite{PhysRevA.103.032211, PhysRevResearch.4.013157, PhysRevB.99.235432, PhysRevLett.119.170602}, but it can be more important for practical applications because of the concern on time cost. As the Carnot cycle achieves the Carnot efficiency only in the limit of infinite period and zero power~\cite{PhysRevE.105.034102, PhysRevLett.106.230602, Polettini_2017, PhysRevE.62.6021, PhysRevE.95.052128, PhysRevLett.110.070603, Nat.Commun.7:11895, PhysRevB.87.165419, PhysRevB.94.121402, PhysRevLett.112.140601, PhysRevLett.114.146801, Sothmann_2014, PhysRevE.98.042112, PhysRevLett.124.110606, PhysRevE.96.062107, PhysRevLett.121.120601, PhysRevE.103.042125, Lee:2017aa, PhysRevE.101.052132}, considering on the maximum power might be more significant than considering on the efficiency in practice. Comparison of the power among various types of engines also lacks discussions so far.

As previous researches have already revealed the impact of the coupling between internal degrees of freedom on the performance of quantum thermal machines~\cite{PhysRevE.104.014149, PhysRevA.102.062422}, it is natural to be curious about its influence on the power of a quantum heat engine. So far, there have been a lot of applications of quantum thermal machines with different kinds of coupling \cite{PhysRevE.96.062120, PhysRevE.90.032102, PhysRevA.102.062419}. Different couplings can play key roles in some quantum thermal machines, such as the minimal two-body quantum absorption refrigerator achieved by $XX$ and $ZZ$-couplings \cite{PhysRevB.104.075442}.

To examine the influence of the coupling between two qubits in the internal system, we here consider an $XX$-coupling \cite{PhysRevA.98.052123,  Naseem_2020, PhysRevB.104.075442} in our double-qubit system. In the present paper, we resolve questions from the viewpoint of gaining a greater power and observe several interesting conclusions. 

We first find out for the single-qubit Otto engine an approximately linear relation between the temperature difference of the heat baths and the energy-level difference of the internal system for the maximum power. We can thereby tune other parameters to achieve the maximum power under the fixed ratio of the heat-bath temperatures. 
We then define four models of the coupled-qubit quantum heat system based on quantum optical two-atom thermal diode \cite{PhysRevE.99.042121}. The two qubits named Q1 and Q2 here can have different energy levels, but for the comparison we make the qubit Q1 which produces the work,  maintain the energy-level change given by the linear relation for the single-qubit Otto engine to obtain the maximum power for a fixed ratio of bath temperatures. 

Utilizing the simulation assisted by a Python quantum tool called QuTip~\cite{qutip1,qutip2}, we numerically observe that the coupling and the positions of attaching the heat baths influence the coupled-qubit system in various ways. 
First, the positions of attaching the heat baths and the coupling affects the difficulties of achieving convergence to a limit cycle for our couple-qubit systems. When each heat bath interacts with the internal system always through one unique qubit, which we will refer to as Model~11 and Model~22, the coupling strength should be stronger for obtaining a limit cycle from an initial state when the energy levels of Q1 become higher. In contrast, when each bath interacts with the coupled-qubit system through different qubits, which we will refer to as Model~12 and Model~21, the cycle converges to a limit cycle quickly. 

Second, we find that all of our models break the maximum-power relation of the single-qubit system and achieve much greater power than the single-qubit one, when we keep the other parameters except the coupling equal to the ones in the single-qubit case. 
With a fixed coupling strength, Model~11, Model~21/Model~12 and Model~22 produce the maximum power from the greatest to the lowest in this order, and the maximum powers of all of them are greater than that of the single-qubit one. However, Model~11 achieves the maximum power only with high energy levels of Q1, and hence the influence of the coupling is not very visible. Besides, Model~11 does not converge to a limit cycle quickly. We thus focus on the other three models for application purposes. For all of these systems, the system efficiency at the maximum power is lower than the Otto efficiency, not being equal to the Otto efficiency as the single-qubit system. In short, the coupling increases the maximum power while it decreases the system efficiency, which is consistent with a trade-off relation between the efficiency and the power~\cite{PhysRevLett.117.190601, PhysRevE.106.024137, PhysRevLett.120.190602, PhysRevE.97.062101, Dechant_2019, Koyuk_2019}. The system efficiency of the maximum power in our schemes is lower than the Curzon-Ahlborn efficiency, which is the efficiency when the Carnot cycle produces the maximum power, whereas the single-qubit system yields a higher system efficiency than the Curzon-Ahlborn efficiency.

This paper is organized as follows. In Sec.~II, we review the model of single-qubit system and explain the method and results of its analysis, particularly from the point of view of the maximum power, focusing on the discovery of a linear relation between the temperatures of the heat baths and the energy levels of the internal qubit. In Sec.~III, we outline our coupled-qubit models, discuss the physical mechanism behind our Otto quantum thermal machines, and define main physical quantities. In Sec.~IV, we explain the dynamics for the interaction between the internal system and the external environment in the process of heat and work exchanges. In Sec.~V, we present the  results and make the comparison of the models. Finally, Sec.~VI is devoted to a summary and conclusions.

\section{Single-qubit System}
In this section, we overview the model and calculation of a single-qubit cycle~\cite{PhysRevResearch.3.023078, PhysRevResearch.5.023066} for later comparison with a double-qubit cycle examined in the following sections. Focusing on the power, we find a linear relation between the temperature difference between the heat baths and the energy-level difference of the internal system at the point of achieving the maximum power, which will be useful for us to come up with schemes of the coupled-qubit heat machine in Sec.~III.

\subsection{Single-qubit Otto Cycle}
\begin{figure}
	\centering
	\includegraphics[scale=0.6]{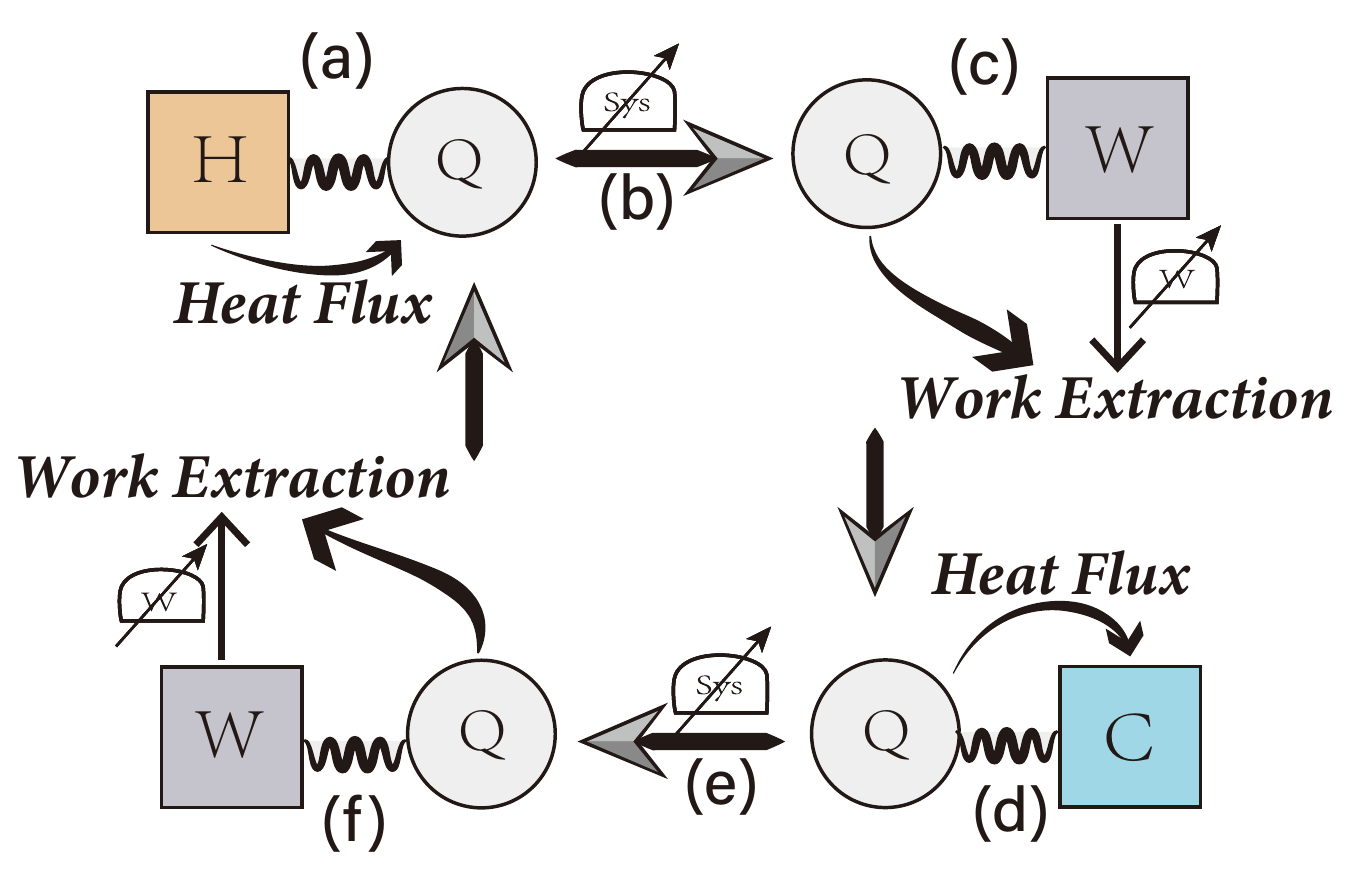}
	\caption{Schematic view of the single-qubit Otto cycle}
	\label{SQ_cycle}
\end{figure}

The most elementary quantum Otto heat engine is composed of one qubit, two heat baths and two work storages going through two ischoric processes and two adiabatic work-production processes~\cite{PhysRevResearch.5.023066, PhysRevA.95.032132}. The single-qubit quantum Otto cycle operates in six steps~(a)--(f) as shown in Fig.~\ref{SQ_cycle}, where we assume that heat and work are positive when they flow from the external environment to the internal system. 
Initially, the qubit of energy gap $\omega_h$ is at the ground state. (a)~When it interacts with the hot bath of temperature $T_h$ for the time duration $t_h$, it is excited, obtaining heat $Q_h$ from the hot bath;
(b)~projection measurement of the qubit is carried out, severing quantum entanglement with the hot bath;
(c)~the energy gap is decreased from $\omega_h$ to $\omega_c$ when the qubit interacts with the work storage, transfering energy $-W_1$ to the storage;
(d)~the qubit of the energy gap $\omega_c$ interacts with the cold bath of temperature $T_c$ for the time duration of $t_c$ and is de-excited, discarding heat $-Q_c$ into the cold bath;
(e)~projection measurement of the qubit is carried out again, severing quantum entanglement with the cold bath;
(f)~the energy gap is increased back from $\omega_c$ to $\omega_h$ when the qubit interacts with the work storage, transfering the energy $W_2$ from the storage. The cycle goes back to the step~(a).
We let the cycle continue until it converges to a limit cycle. In numerical simulations we use the convergence criteria based on the energy conservation:
\begin{align}
	\label{EnergyConvergence}
	\begin{split}
	\triangle E  &:= |Q_h+Q_c+W_1+W_2| \\&\leq \mathrm{min}[|Q_h|, |Q_c|, |W_1|, |W_2|] \times 10^{-2},
	\end{split}
\end{align}
where we define $Q_h$, $Q_c$, $W_1$ and $W_2$ below in Eqs.~(\ref{SQ_Qh})--(\ref{SQ_W2}).
We let $N$ denote the number of iterations before the convergence.

The steps~(a) and (d) are isochoric processes, whereas the steps~(c) and (f) correspond to the adiabatic expansion and compression of the classical Otto cycle, respectively. The whole Hamiltonians of the first and second isochoric processes are given by
\begin{align}
	&H_{\mathrm{iso}}^\alpha = H_S^\alpha+H_B^\alpha+H_{\mathrm{int}}^\alpha,
\end{align}
where $\alpha=h,c$ denote the instance of the interaction with the hot and cold baths, respectively. The system Hamiltonians of the first and second halves of the cycle are respectively given by
\begin{align}
	H_S^\alpha &=
	\begin{pmatrix}
		0 & 0  \\
		0 & \omega_\alpha  \\
	\end{pmatrix},
\end{align}
with $\alpha=h,c$, where $\omega_h>\omega_c>0$, and we put $\hbar$ to unity here and hereafter. We use $\omega_c$ as the energy unit and $1/\omega_c$ as the time unit. For the isochoric processes in the steps~(a) and (d), we employ bosonic heat baths whose Hamiltonian is given by
\begin{align}
	\label{BathHamiltonian}
	H_B^\alpha = \sum_{\mu}\epsilon_{\mu, \alpha}\hat{a}_{\mu, \alpha}^\dagger \hat{a}_{\mu, \alpha},
\end{align}
where $\hat{a}_{\mu,\alpha}^\dagger$ and $\hat{a}_{\mu,\alpha}$ are the creation and annihilation operators of the mode $k$ of the bath $\alpha$. The contact Hamiltonian between the single-qubit system and each bath $\alpha=h,c$ is 
\begin{align}
	H_{\mathrm{int}}^{\alpha} =  \sum_{\mu}g_{\mu, \alpha} \sigma_S^x (\hat{a}_{\mu, \alpha}^\dagger+\hat{a}_{\mu, \alpha}),
\end{align}
where $\sigma_S^x$ represents the $x$ component of the Pauli matrices of the qubit, $g_{\mu,\alpha}$ is the coupling strength between the internal qubit and the mode $\mu$ of the bath $\alpha$. 

As detailed in Sec~II.~B, we analyze the time evolution of the single-qubit system under the interaction with heat baths by means of the standard master equation. As we describe details in Sec~II.~C, on the other hand, we perform the work production processes as an indirect-measurement model \cite{PhysRevResearch.5.023066, PhysRevA.95.032132} using quantum measurement theory; we measure the energy increase and decrease of the work storage after interaction between the work storage and the system. Since these work-production processes do not change the state of the system, we assume that it takes a negligible time of extracting work. 

We define $t=0$ as the starting point of step~(a) after the Otto cycle achieves the convergence; the interaction between the internal system and the hot bath leads the state of the internal system to change from $\rho(0)$ to $\rho(t_h)$, and the interaction between the internal system and the cold bath lets the state of the internal system evolve from $\rho(t_h)$ to $\rho(t_h+t_c)$. We then follow the standard definition of heat transfer:
\begin{align}
	\label{SQ_Qh}
	&Q_h = \tr[H_S^h(\rho(t_h)-\rho(0))], \\
	\label{SQ_Qc}
	&Q_c = \tr[H_S^c(\rho(t_c+t_h)-\rho(t_h))].
\end{align}
On the other hand, the work production is typically defined as
\begin{align}
	&W = W_1+W_2,\\
	\label{SQ_W1}
	&W_1 = \tr[\rho(t_h)(H_S^c-H_S^h)], \\
	\label{SQ_W2}
	&W_2 = \tr[\rho(t_h+t_c)(H_S^h-H_S^c)].
\end{align}
We will reconsider the definition of work in Sec.~II.~C using the indirect measurement theory, but the bottom line will be the same.

There are three possible types of thermal machines depending on the signs of heat and work. When the system makes the heat flow from the hot bath at a higher temperature to the cold bath at a lower temperature obtaining work from the environment, the quantum Otto thermal machine operates as a heater. If the thermal machine extracts the heat from the cold bath and makes it flow into the hot bath, there must be work given by the external environment to the internal system because of the second law of thermodynamics, and it is a refrigerator. The last one is a quantum heat engine, which is the focus of the present paper. In the case of the quantum heat engine, the heat flows from the hot bath to the cold bath, which is similar to the heat exchange of the heater, but the work is produced by the internal system to the external environment. 

In other words, the definitions of these different thermal machines are given as follows; (a)~for a heater, $Q_h>0, Q_c<0, W>0$; (b)~for a cooler, $Q_h<0, Q_c>0, W>0$; and (c)~for an engine, $Q_h>0, Q_c<0, W<0$. 
The heater's coefficienct of the performance (HCOP) for the case~(a) and the cooler's coefficienct of the performance (CCOP) for the case~(b) as well as the power and the efficiency for the case~(c) are defined by 
\begin{align}
	&\mathrm{HCOP} = -\frac{Q_c}{W_1+W_2},\\
	&\mathrm{CCOP} = \frac{Q_c}{W_1+W_2},\\
	\label{P}
	&P = -\frac{(W_1+W_2)}{t_h+t_c},\\
	\label{Eff}
	&\eta = -\frac{W_1+W_2}{Q_h}.
\end{align}

\subsection{Standard Master Equation}
For the single-qubit quantum Otto cycle \cite{PhysRevE.76.031105}, we employ the standard Gorini-Kossakowski-Sudarshan-Lindblad (GKSL)~\cite{Hofer_2017, doi:10.1142/S1230161217400108} master equation under the Born-Markov approximation and weak-coupling approximation to simulate the single-qubit machine numerically using the Python toolbox Qutip~\cite{qutip1,qutip2}. 

In the process of interaction between the qubit and each heat bath $\alpha=h,c$ at the steps~(a) and (d), respectively, we simulate
\begin{align}
	\frac{d\rho}{dt}=-i[H_S^\alpha,\rho]+\hat{L}_\alpha\rho,
\end{align}
where the Liouville superoperator $\hat{L}_\alpha$ $(\alpha=h, c)$ is given by
\begin{align}
	\hat{L}_\alpha\rho=(G_\alpha(\omega_\alpha)\hat{D}[\sigma^-]+G_\alpha(-\omega_\alpha)\hat{D}[\sigma^+])\rho
\end{align} 
with the Lindblad dissipators
\begin{align}
	\label{Lindblad}
	\hat{D}[\hat{o}]\rho=\frac{1}{2}(2\hat{o}\rho\hat{o}^\dagger-\hat{o}^\dagger\hat{o}\rho-\rho\hat{o}^\dagger\hat{o})
\end{align}
and the spectral response functions of the thermal baths
\begin{equation}
	\label{SRF}
	G_\alpha(\omega)= \gamma_\alpha(\omega)(1+\bar{n}_\alpha(\omega))+\gamma_\alpha(-\omega)\bar{n}_\alpha(\omega)
\end{equation}
for heat baths $\alpha=h, c$, where $n_\alpha(\omega)$ is the Bose-Einstein distribution at temperature $T$ given by
\begin{align}
	\label{bath}
	\bar{n}_\alpha(\omega)=\frac{1}{e^{\omega / k_bT}-1}
\end{align}
with the zero chemical potential. The function $\gamma_\alpha(\omega)$ is the energy damping rate \cite{RevModPhys.59.1} for the interaction between the qubit and the bath $\alpha$, given by
\begin{align}
\label{DR}
\begin{split}
&\gamma_\alpha(\omega)=\\
&\begin{cases}
	2\pi\sum_{\mu}g_{\mu,\alpha}^2\delta(\omega-\omega_{\mu,\alpha})=2\pi J_\alpha(\omega),&\text{for $\omega>0$},\\
	0, &\text{for $\omega\leq0$},
\end{cases}
\end{split}
\end{align}
where $g_{\mu,\alpha}$ is the interaction strength between the qubit and the $\mu$th oscillator of the bath $\alpha$, $\omega_{\mu,\alpha}$ is the frequency of the oscillator, and the function $J_\alpha(\omega)$ is given by
\begin{align}
	\label{SDF}
	J_\alpha(\omega) &=\kappa_\alpha\frac{\omega^{s}}{\omega_{\mathrm{ct}}^{1-s}}\exp(-\frac{\omega}{\omega_{\mathrm{ct}}})
\end{align}
with the cut-off frequency $\omega_{\mathrm{ct}}$ and the transition rates $\kappa_\alpha$ of heat bath $\alpha$. In the present work, we consider the Ohmic spectral density for each bath with $s = 1$.

\subsection{Work Extraction Process}
As we describe above, we perform the working production processes as an indirect-measurement model \cite{PhysRevResearch.5.023066, PhysRevA.95.032132} using quantum-measurement theory, by measuring the energy increase of the work storage after interaction between the work storage and the system. 

In order to keep track of the variation of the qubit’s Hamiltonian in the interaction process, we introduce a clock as an additional degree of freedom. The Hamiltonian $H_{\mathrm{SW}}$ of the total system thus consists of the qubit $H_S^\alpha$ ($\alpha=h, c$), the clock $C$ and the work storage $H_W$ in the first process of the work production in the step~(c) of Fig.~\ref{SQ_cycle}:
\begin{align}
	\begin{split}
	H_{\mathrm{SW}} =&H_S^h\otimes\ket{0}_C\bra{0}\otimes\mathbb{I}_W \\&+H_S^c\otimes\ket{1}_C\bra{1}\otimes\mathbb{I}_W +H_W, 
	\end{split}\\
	H_W = &\mathbb{I}_S\otimes\mathbb{I}_C\otimes(\omega_c-\omega_h)\ket{1}_W\bra{1}.
\end{align}
We set the state of the total system before the work extraction to
\begin{align}
\begin{split}
	\rho_{\mathrm{SW}}^\mathrm{i} = &\left(\rho_{00}\ket{0}_S\bra{0} +\rho_{11}\ket{1}_S\bra{1}\right) \\
	&\otimes \ket{0}_C\bra{0} \otimes \ket{0}_W\bra{0},
\end{split}
\end{align}
while after the work extraction, the clock flips, but the state of the work storage changes only for the excited state of the internal single-qubit system. Therefore, the state of the total system becomes
\begin{align}
\begin{split}
	\rho_{\mathrm{SW}}^\mathrm{f} = &\rho_{00}\ket{0}_S\bra{0} \otimes\ket{1}_C\bra{1} \otimes\ket{0}_W\bra{0}\\ &+\rho_{11}\ket{1}_S\bra{1} \otimes\ket{1}_C\bra{1} \otimes\ket{1}_W\bra{1}.
\end{split}
\end{align}
Naturally, we can define a quenched unitary transformation for the process:
\begin{align}
\begin{split}
	U_{\mathrm{SW}} =&\ket{0}_S\bra{0} \otimes (\ket{0}_C\bra{1}+\ket{1}_C\bra{0})\otimes\mathbb{I}_W \\
	&+\ket{1}_S\bra{1} \otimes (\ket{00}_{CW}\bra{11}+\ket{11}_{CW}\bra{00}\\ &+\ket{01}_{CW}\bra{01}+\ket{10}_{CW}\bra{10}).
\end{split}
\end{align}
This unitary operation commutes with the total Hamiltonian $H_{\mathrm{SW}}$, hence satisfying the energy conservation law. 

Performing the projective measurement to the work storage, we can observe the probabilities of work state without destroying the state of the internal system. Let us set the projection operators as follows:
\begin{align}
	\label{P0}
	P_0 = \ket{0}_W\bra{0}, \\
	\label{P1}
	P_1 = \ket{1}_W\bra{1}. 
\end{align}
Then the work extraction through the observation is calculated by
\begin{align}
	W_1 &= \tr[H_W(\rho_{11}\ket{1}_W\bra{1}+\rho_{00}\ket{0}_W\bra{0})] \\ &=(\omega_h-\omega_c)\rho_{11}.
\end{align} 
Obviously, it is equal to the result we would obtain by the elementary definition of work in Eq.~(\ref{SQ_W1}) for the particular choice of the projection operators~(\ref{P0}) and (\ref{P1}). We can similarly derive Eq.~(\ref{SQ_W2}) employing the indirect-measurement scheme.

\subsection{Numerical Results}
\begin{figure*}
	\centering
	\subcaptionbox[subcaption3]{Power\label{SQ:power}}[0.4\textwidth]{
		\includegraphics[width=0.4\textwidth]{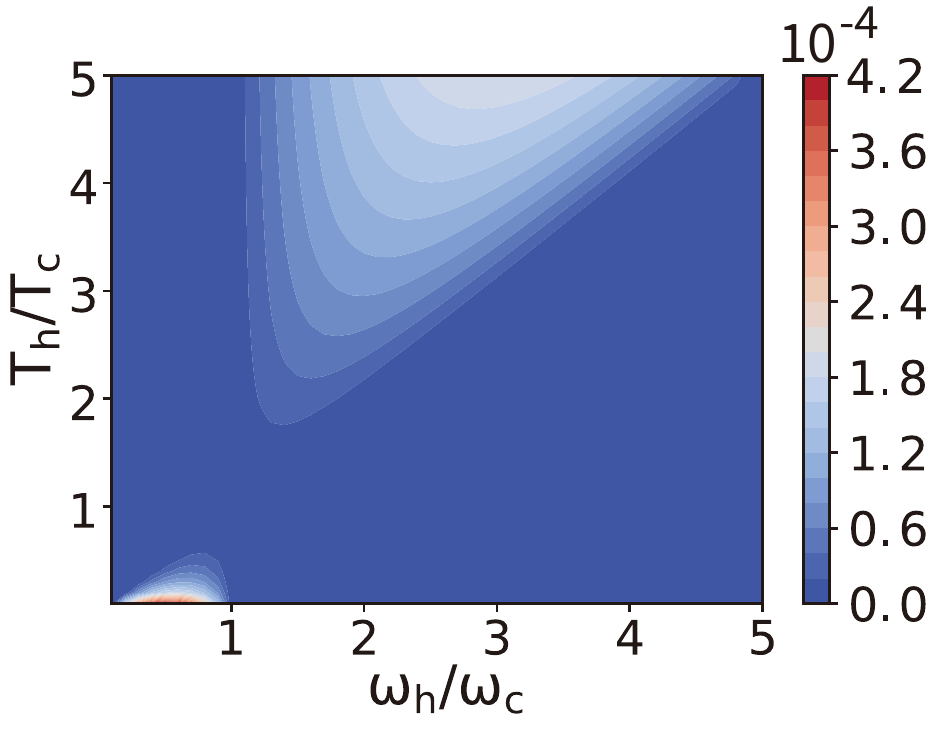}}
		\hspace{0.05\textwidth}
	\subcaptionbox[subcaption1]{Efficiency\label{SQ:efficiency}}[0.4\textwidth]{
		\includegraphics[width=0.4\textwidth]{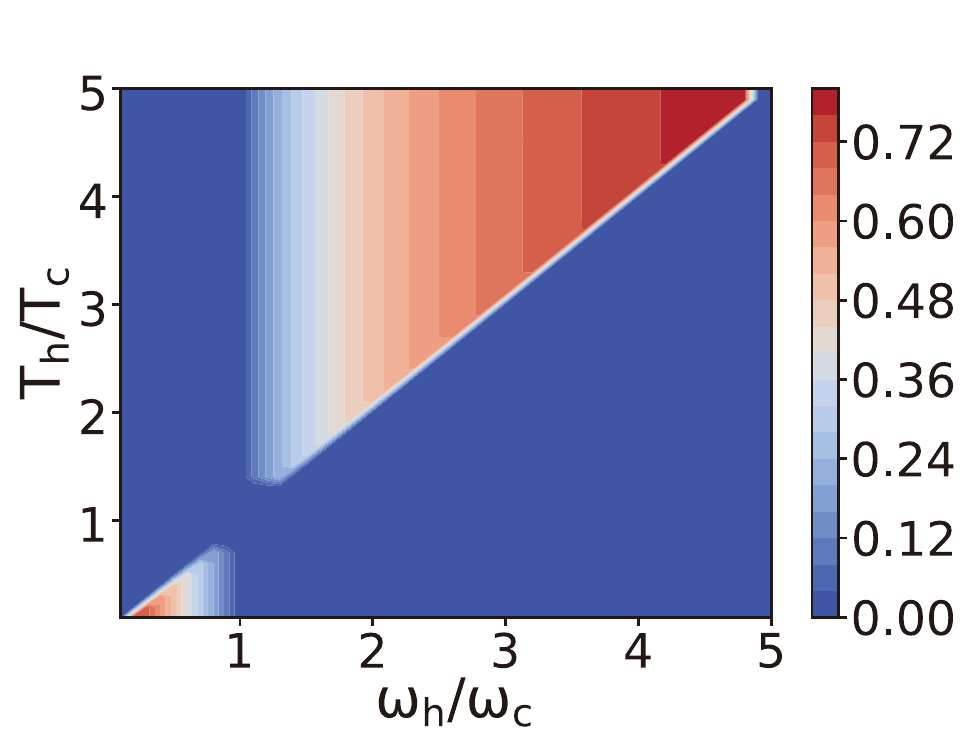}}
		\\[\baselineskip]
	\subcaptionbox[subcaption4]{HCOP\label{SQ:HCOP}}[0.4\textwidth]{
		\includegraphics[width=0.4\textwidth]{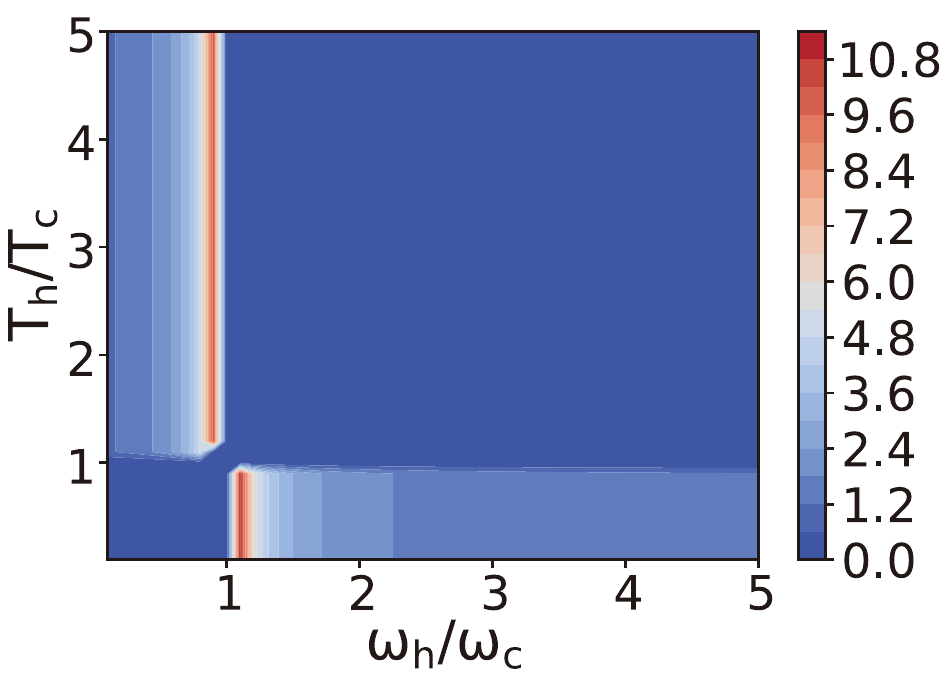}}
		\hspace{0.05\textwidth}
	\subcaptionbox[subcaption5]{CCOP\label{SQ:CCOP}}[0.4\textwidth]{
		\includegraphics[width=0.4\textwidth]{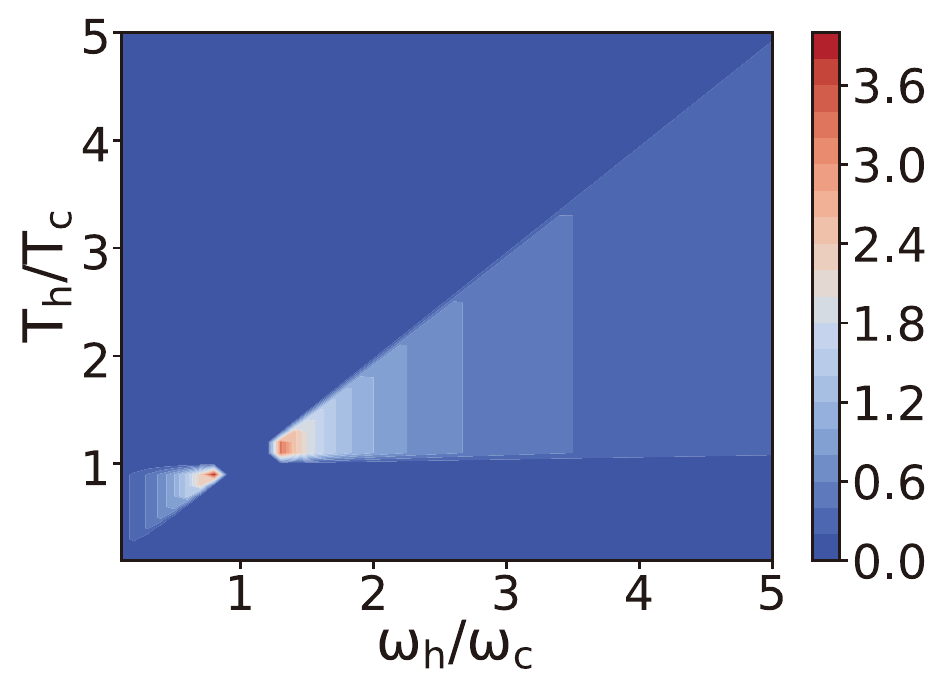}}
	\caption{(a)~Power, (b)~efficiency, (c)~HCOP and (d)~CCOP for the single-qubit Otto machine as functions of the energy levels $\omega_h/\omega_c$ and the heat-bath temperatures $T_h/T_c$. Parameters: energy unit $\omega_c=1$; the temperature of the cold bath $T_c = 5$; the transition rate $\kappa_h = \kappa_c = 0.005$; the time durations $t_h = t_c = 50$.}
	\label{SQ_density}
\end{figure*}

For numerical simulation of the single-qubit machine, we set the lower excited energy level to $\omega_c=1$, which is also the energy unit, and fix the temperature of the cold bath to $T_c=5$ for later comparison. We also fix the transition rate to $\kappa_h = \kappa_c=0.005$ and the time durations of the interaction between the internal single-qubit system and each heat bath to $t_h=t_c=50$. We then pursue the dependence of physical quantities,  varying the higher energy level $\omega_h$ and the hot-bath temperatures $T_h$. We observe that the single-qubit Otto cycle operates as different thermal machines under diverse circumstances with different parameters. As shown in Fig.~\ref{SQ_density}, the single-qubit machine acts as an engine when $T_h/T_c>\omega_h/\omega_c>1$ or $T_h/T_c<\omega_h/\omega_c<1$, as a heater when $\omega_h/\omega_c<1$ and $T_h/T_c>1$ or $\omega_h/\omega_c>1$ and $T_h/T_c<1$, and as a cooler when $1<T_h/T_c<\omega_h/\omega_c$ or $1>T_h/T_c>\omega_h/\omega_c$. We hereafter focus on the first case.

\begin{figure*}
	\centering
	\subcaptionbox[subcaption1]{Power\label{SQ:P_kw}}[0.35\textwidth]{
		\includegraphics[scale=1.0]{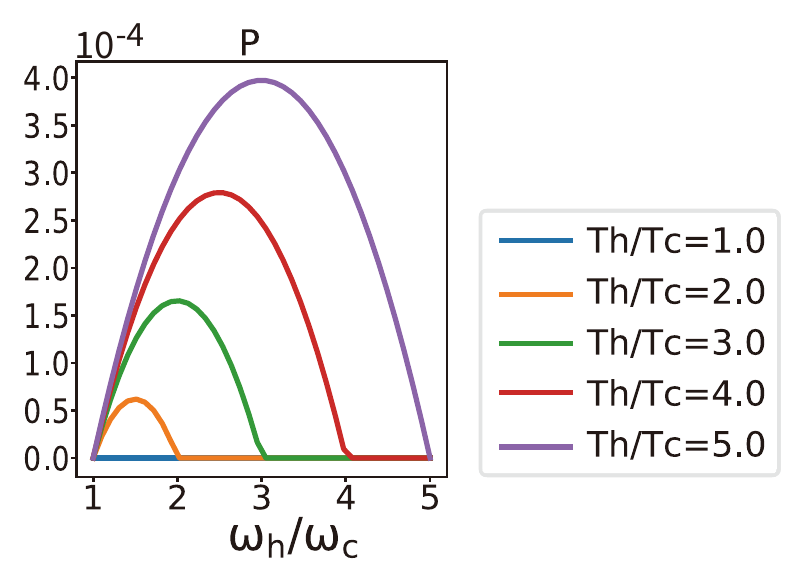}}
	\subcaptionbox[subcaption2]{Efficiency\label{SQ:Eff_kw}}[0.45\textwidth]{
		\includegraphics[scale=1.0]{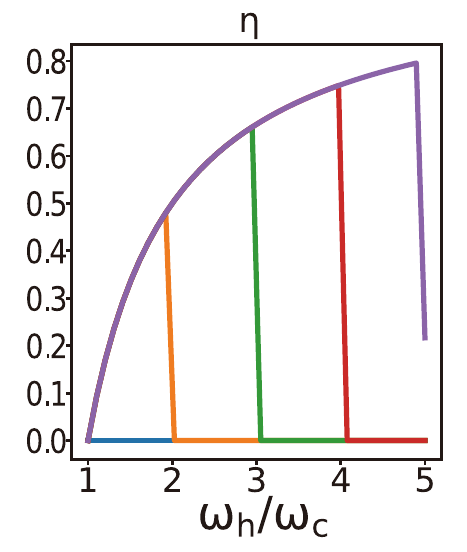}}\par
	\caption{Dependence of (a)~power $P$ and (b)~efficiency $\eta$ of the single-qubit engine on the energy levels $\omega_h/\omega_c$. Parameters: energy unit $\omega_c = 1$, the temperature of cold bath $T_c = 5$; the transition rate $\kappa_h = \kappa_c = 0.005$; the time durations $t_h = t_c = 50$}
	\label{SQ_Pkw}
\end{figure*}

\begin{figure*}
	\centering
	\subcaptionbox[subcaption1]{$P_m$\label{SQ:Pm}}[0.31\textwidth]{
		\includegraphics[scale=0.83]{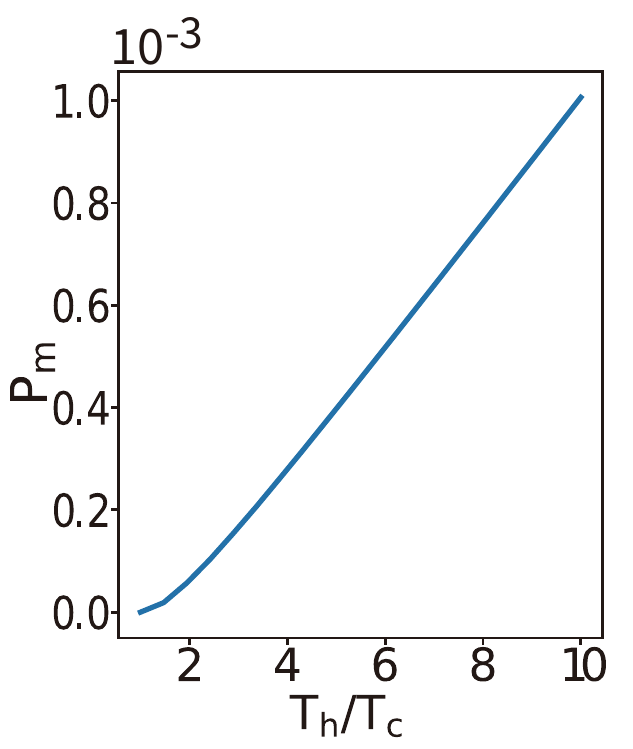}}
	\subcaptionbox[subcaption2]{$(\omega_h/\omega_c)_{P_m}$\label{SQ:kw}}[0.31\textwidth]{
		\includegraphics[scale=0.83]{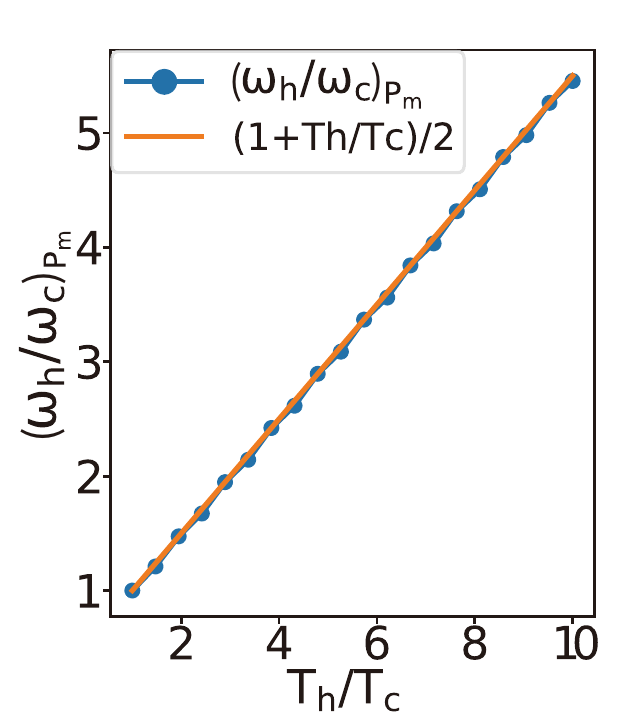}}
	\subcaptionbox[subcaption3]{$\eta_{P_m}$\label{SQ:Eff}}[0.31\textwidth]{
		\includegraphics[scale=0.83]{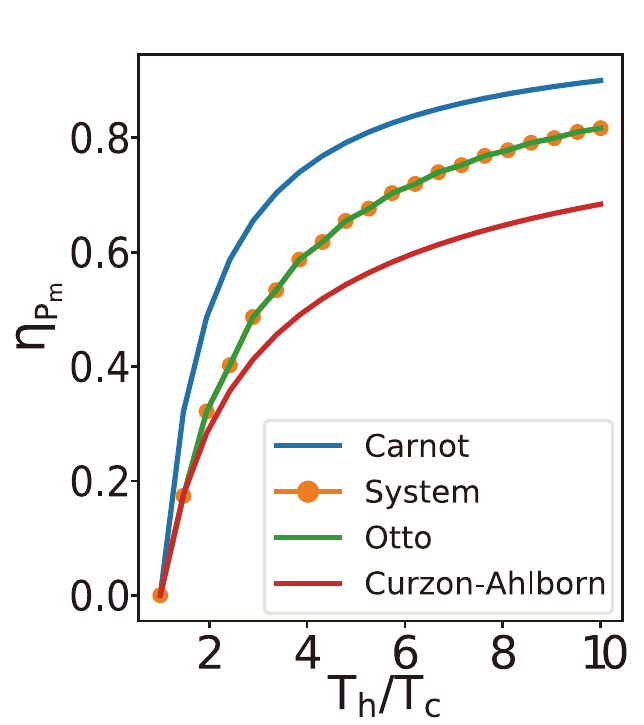}}
	\caption{Dependence of (a)~the maximum power $P_m$, (b)~the energy levels $(\omega_h/\omega_c)_{P_m}$ and (c)~the efficiency $\eta_{P_m}$ on the heat-bath temperatures $T_h/T_c$ for single-qubit Otto engine. Parameters: energy unit $\omega_c=1$, temperature of cold bath $T_c = 5$, transition rate: $\kappa_h = \kappa_c = 0.005$, time durations: $t_h = t_c = 50$}
	\label{SQ_PmIndex}
\end{figure*}

When the single-qubit Otto system runs as a quantum heat engine, its efficiency $\eta$ and the power $P$ behave as shown in Fig.~\ref{SQ:power} and Fig.~\ref{SQ:efficiency}, respectively, depending on the ratio of the energy gaps and the temperatures.  
As shown in Fig.~\ref{SQ:power}, the power has a shape of a semicircular cone depending on the ratio of the heat baths' temperatures $T_h/T_c$ and the system's energy levels $\omega_h/\omega_c$. Therefore, it is easy to find a peak of the power as we scan the ratio of energy levels with the temperatures fixed; as shown explicitly in Fig.~\ref{SQ:P_kw}, the power of the single-qubit Otto engine always has a unique peak point as a function of the ratio of the system's energy levels for a fixed ratio of the heat baths' temperatures. We define this peak as the maximum power of the single-qubit Otto engine that we are interested in:
\begin{align}
	P_m = P_m(T_h/T_c) := \mathop{\max}_{\omega_h/\omega_c} P(T_h/T_c,\omega_h/\omega_c).
\end{align} 
We also define the ratio of the energy levels and the efficiency at the parameter point of the maximum power: 
\begin{align}
	&(\omega_h/\omega_c)_{P_m}: = \mathop{\mathrm{argmax}}_{\omega_h/\omega_c}P(T_h/T_c,\omega_h/\omega_c),\\
	&\eta_{P_m} := \eta(T_h/T_c, (\omega_h/\omega_c)_{P_m}),
\end{align}
which are presented in Fig.~\ref{SQ_PmIndex}. 

At a fixed temperature of the cold bath, the maximum power increases as the hot-bath temperature grows. We notice in Fig.~\ref{SQ:kw} that there is a linear relation between $(\omega_h/\omega_c)_{P_m}$ and $T_h/T_c$ of the form
\begin{align}
	\label{MPR1}
	\left(\frac{\omega_h}{\omega_c}\right)_{P_m}=\frac{1}{2}\left(1+\frac{T_h}{T_c}\right).
\end{align} 
In other words, the power becomes maximum when $\omega_h$ is set to
\begin{align}
	\label{MPR2}
	&\Omega_h = \frac{\omega_c}{2}\left(1+\frac{T_h}{T_c}\right),
\end{align}
and hence the energy-level change of the qubit is set to
\begin{align}
	\label{MPR3}
	&\Delta\omega = \Omega_h-\omega_c = \frac{\omega_c}{2}\left(\frac{T_h}{T_c}-1\right).
\end{align}

The four efficiencies, namely the system efficiency, the Otto efficiency, the Carnot efficiency and the Curzon-Ahlborn efficiency are plotted in Fig.~\ref{SQ:Eff}, at the maximum power point of the single-qubit Otto engine. The efficiency of the single-qubit engine is equal to the quantum Otto efficiency~\cite{PhysRevResearch.3.023078}:
\begin{align}
	\label{Otto}
	\eta = \eta_{\mathrm{Otto}} = 1-\frac{\omega_c}{\omega_h}.
\end{align}
The Curzon-Ahlborn efficiency is the efficiency when the Carnot cycle achieves the maximum power. The Otto efficiency at the maximum power is greater than the Curzon-Ahlborn efficiency, which means that the efficiency at the maximum power point in the present single-qubit Otto engine is greater than the Carnot-engine one. 

\section{Coupled-qubit Model}
In this section, we propose the same quantum Otto cycle but in which the working medium is composed of two qubits with the $XX$-coupling. In Sec.~III.~A, based on the quantum Otto cycle and the maximum-power relations~(\ref{MPR1})--(\ref{MPR3}) of the single-qubit Otto system given in Sec.~II, we define four models of the coupled-qubit engine in which each bath contacts each qubit. In Sec.~III.~B, we present the Hamiltonians and physical quantities of our coupled-qubit Otto engines.

\subsection{Four Models}
We now consider the Otto cycle comprised of an internal system of two qubits coupled with an $XX$-coupling and four environment components including two heat baths and two work storages. Initially, both the first qubit Q1 and the second qubit Q2 are at their ground states. The operation protocol is achieved by the following six steps:
(a)~the system has a contact with the hot bath at $\mathrm{Q}\mathit{H}\text{(=Q1 or Q2)}$ and get excited, leading to heat transfer of $Q_h$ from the hot bath to the system; 
(b)~the projection measurement is carried out to severe the quantum entanglement between the system and the hot bath;
(c)~Q1 interacts with the work storage, and the Hamiltonian of Q1 is updated from $H_S^h$ to $H_S^c$, leading to work production $-W_1$;
(d)~the system has a contact with the cold bath at $\mathrm{Q}\mathit{C}\text{(=Q1 or Q2)}$ and get de-excited, leading to heat transfer of $-Q_c$ from the system to the cold bath;
(e)~the projection measurement is carried out again to severe the quantum entanglement between the system and the cold bath;
(f)~Q1 interacts with the work storage, and the Hamiltonian of Q1 is changed from $H_S^h$ back to $H_S^c$, leading to work intake $W_2$. Then the cycle completes and comes back to the step~(a). We let the cycle continue until it converges to a steady limit cycle. For the convergence criteria in numerical simulations, we use the same condition as Eq.~(\ref{EnergyConvergence}), and we again let $N$ denote the number of the Otto cycle iterations before the convergence. 

\begin{figure*}
	\centering
	\subcaptionbox[subcaption1]{Model~11\label{M11_cycle}}[0.49\textwidth]{
		\includegraphics[scale=0.75]{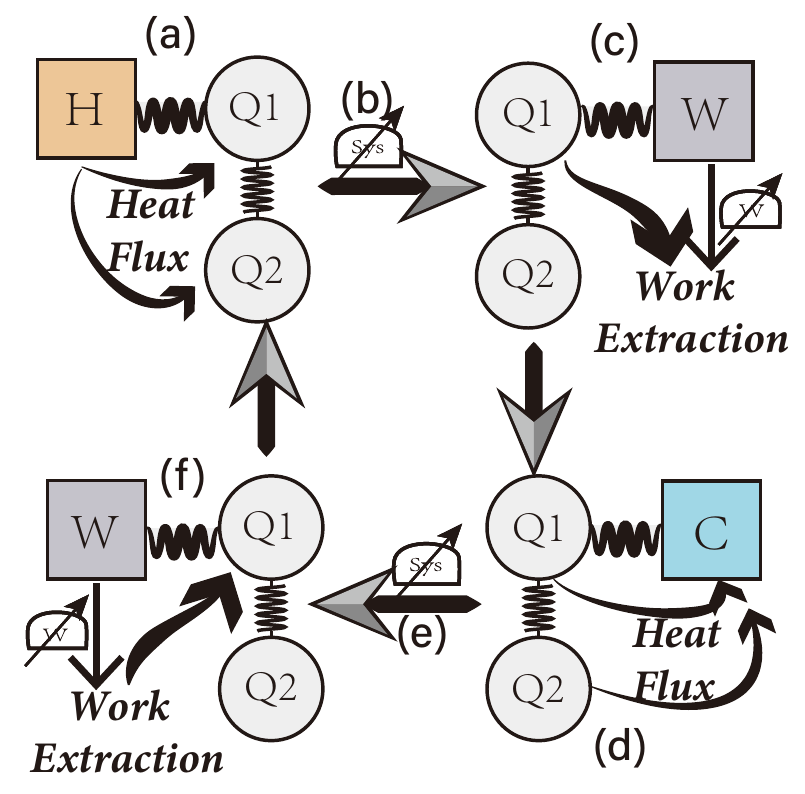}}
	\subcaptionbox[subcaption2]{Model~12\label{M12_cycle}}[0.49\textwidth]{
		\includegraphics[scale=0.75]{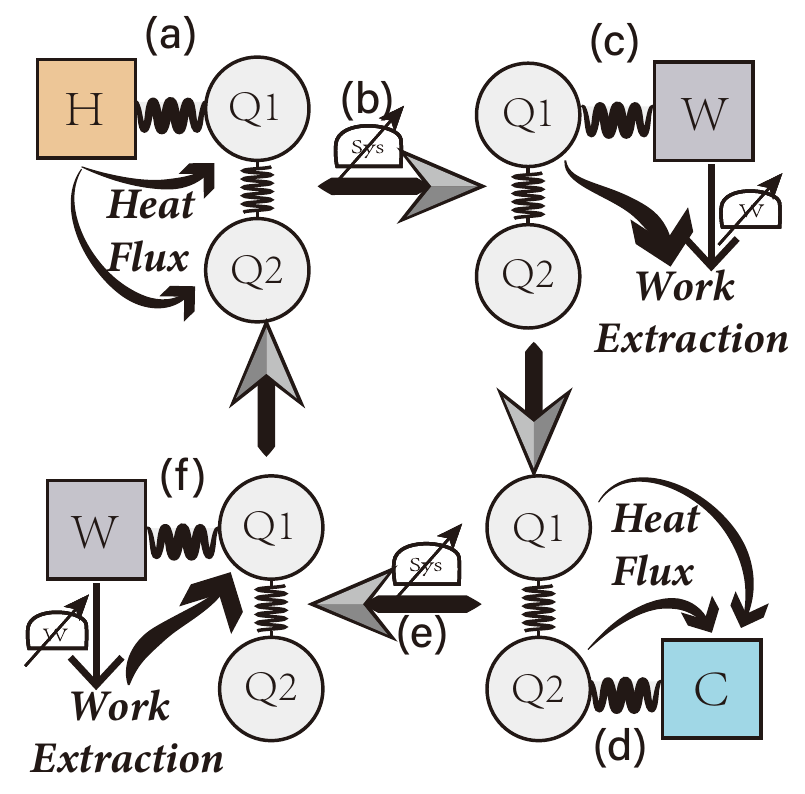}}
	\subcaptionbox[subcaption3]{Model~21\label{M21_cycle}}[0.49\textwidth]{
		\includegraphics[scale=0.75]{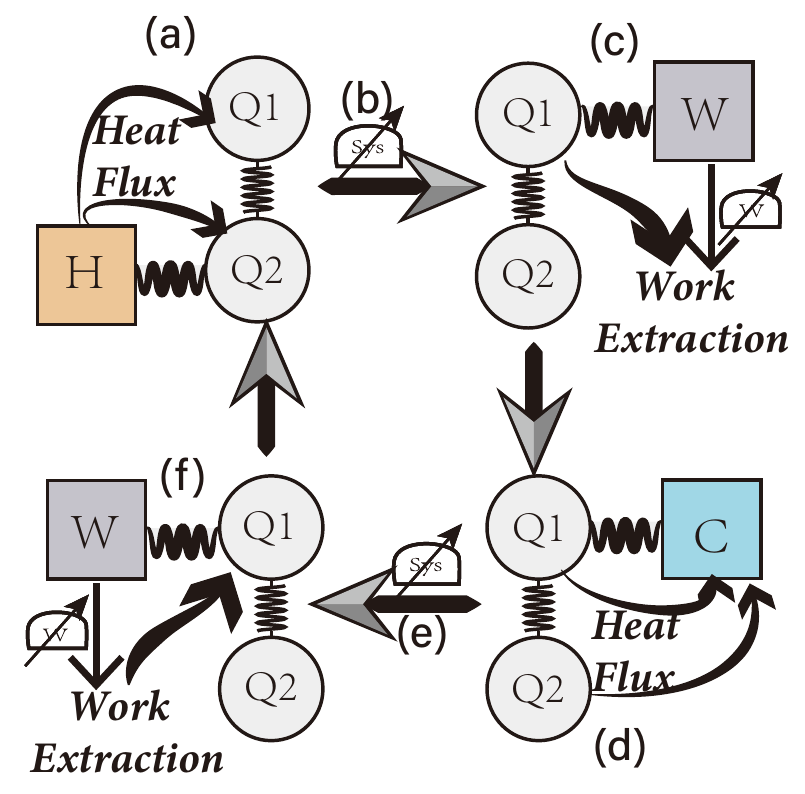}}
	\subcaptionbox[subcaption4]{Model~22\label{M22_cycle}}[0.49\textwidth]{
		\includegraphics[scale=0.75]{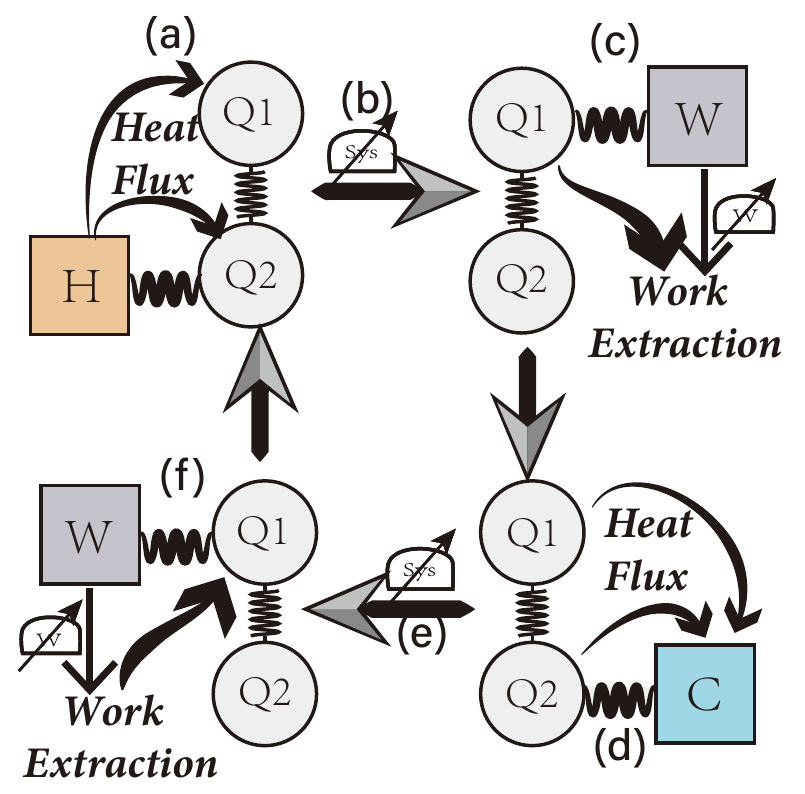}}
	\caption{Schematic views of (a)~Model~11, (b)~Model~12, (c)~Model~21 and (d)~Model~22.}
	\label{cycles}
\end{figure*}

The symbols $\mathrm{Q}\mathit{H}$ and $\mathrm{Q}\mathit{C}$ denote the qubits of the internal systems with $H=1,2$ and $C=1,2$, so that we have four possible schemes of our model, namely Model~11, Model~12, Model~21 and Model~22 as shown in Table~1. Schematic views of the four models are shown in Fig.~\ref{cycles}.
\begin{table}
\begin{tabular}{| c | c | c |}
	\hline 
	$\mathrm{Q}\mathit{H}\backslash\mathrm{Q}\mathit{C}$ & Q1 & Q2 \\ 
	\hline 
	Q1 & Model~11 & Model~12 \\ 
	\hline 
	Q2 & Model~21 & Model~22 \\
	\hline   
\end{tabular}
\caption{Four schemes of our coupled-qubit Otto engines.}
\label{models}
\end{table}

\subsection{Hamiltonians}
For the Otto cycle defined in Sec.~III.~A, the whole Hamiltonians of each isochoric process is given by
\begin{align}
	&H_{\mathrm{iso}}^\alpha = H_S^\alpha+H_B^\alpha+H_\mathrm{int}^{q,\alpha},
\end{align}
where $\alpha=h, c$ denote the case of contact with hot and cold baths, respectively. In the system Hamiltonian
\begin{align}
	\label{DQ_H}
	H_S^\alpha =&H_1^\alpha + H_2 +H_{\mathrm{cp}},
\end{align}
we make only Q1's Hamiltonians $H_1^\alpha (\alpha=h, c)$ change in the process of the work extraction:
\begin{align}
	H_1^\alpha = \omega_1^\alpha\frac{\mathbb{I}_1-\sigma_1^z}{2}\otimes\mathbb{I}_2
\end{align}
with $\alpha=h,c$, where $\mathbb{I}_1$ and $\mathbb{I}_2$ denote the identity operators for the spaces of Q1 and Q2, respectively. Meanwhile, Q2's Hamiltonian $H_2$ and the coupling Hamiltonian $H_{\mathrm{cp}}$ are fixed:
\begin{align}
	H_2 &= \omega_2\mathbb{I}_1\otimes\frac{\mathbb{I}_{2}-\sigma_2^z}{2}, \\
	H_{\mathrm{cp}} &= \frac{g}{2}(\sigma_1^x\sigma_2^x+\sigma_1^y\sigma_2^y) \\&=g(\sigma_1^+\sigma_2^- +\sigma_1^-\sigma_2^+).
\end{align}
Therefore, the system Hamiltonian~(\ref{DQ_H}) is given by
\begin{align}
 	H_S^\alpha
	=& \begin{pmatrix}
		0 & 0 & 0 & 0 \\
		0 & \omega_2 & g & 0 \\
		0 & g & \omega_1^\alpha & 0 \\
		0 & 0 & 0 & \omega_2+\omega_1^\alpha \\
	\end{pmatrix},
\end{align}
under the representation bases $\ket{\mathrm{Q1,Q2}}=(\ket{\downarrow\downarrow}$, $\ket{\downarrow\uparrow}$, $\ket{\uparrow\downarrow}$ and $\ket{\uparrow\uparrow}$) in this order. 
For the heat-bath Hamiltonians $H_\mathrm{int}^{q,\alpha}$, we employ the same bosonic ones as the single-qubit Otto engine given by Eq.~(\ref{BathHamiltonian}).
The interaction Hamiltonians $H_{q,\alpha}$ between either of the system qubits Q1 and Q2 and the heat baths $\alpha (=h,c)$ is given by
\begin{align}
	H_\mathrm{int}^{q,\alpha} &= \sum_{\mu}V_{\mu,\alpha}\sigma_q^x(\hat{a}_{\mu, \alpha}^\dagger+\hat{a}_{\mu, \alpha}),
\end{align}
where $\sigma_q^k$ with $k=x,y,z$ and $q=1$ or $2$ denote the Pauli matrices for the spaces of Q1 and Q2, respectively. 
\begin{figure}
	\centering
	\includegraphics[scale=0.63]{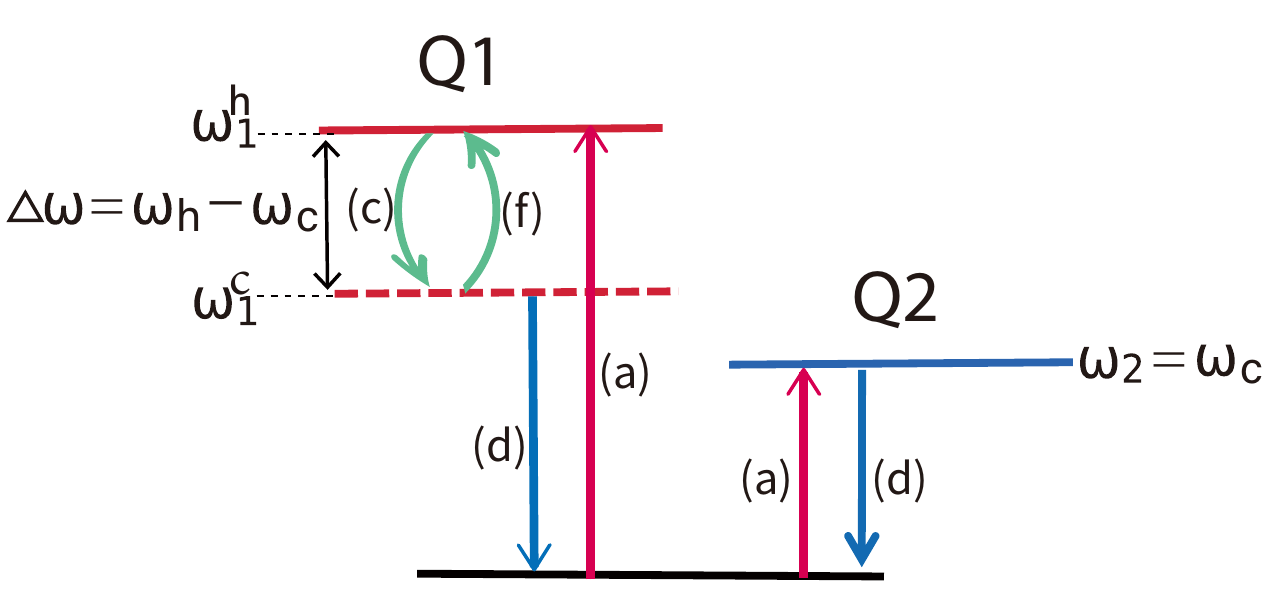}
	\caption{Schematic view of the coupled-qubit Otto engine's energy spectrum.}
	\label{DQ_E}
\end{figure}

For $g=0$, Model~11 should reduce to the single-qubit Otto cycle and the other three models cannot operate successfully. Therefore, by comparing our coupled-qubit models to the single-qubit engine, we examine whether the $XX$-coupling contributes to a greater power and analyze the four models for better applications in various situations. 

In order to compare the single-qubit and coupled-qubit systems on an equal footing, we consider the four models of the coupled-qubit system with the same energy-level change as the case of the maximum power of the single-qubit engine, following the maximum-power relation~(\ref{MPR1})--(\ref{MPR3}) in Sec.~II.~D. Figure~\ref{DQ_E} shows the change of the system at each step of the Otto cycle. Q1's energy levels $\omega_1^\alpha$ in $H_S^\alpha$ can be different from the excited energy $\omega_\alpha$ ($\alpha=h, c$) of the single-qubit system, but for comparison we set the level change $\omega_1^h-\omega_1^c$ equal to $\Delta\omega$ in Eq.~(\ref{MPR3}) of the single-qubit case. 
In other words, we set the energy-level change of Q1 of our coupled-qubit system as follows:
\begin{align}
	& \omega_1^h-\omega_1^c = \Delta\omega =\frac{\omega_c}{2} \left(\frac{T_h}{T_c}-1 \right).
\end{align}
On the other hand, we set the excited energy level $\omega_2$ of Q2 always equal to the lower excited energy $\omega_c$ of the single-qubit Otto engine, which is also the energy unit in the present section:
\begin{align}
	\omega_2 = \omega_c.
\end{align}
In Fig.~\ref{DQ_E}, the red arrows show that the internal system gets excited by the interaction between the internal system and the hot bath, the blue arrows show that the internal system gets de-excited by the interaction between the internal system and the cold bath, and the green arrows show change of the energy levels in the processes of producing and intaking the work.

\section{Dynamics}
For interaction between systems and heat baths, we can address the question as to which of the standard and global Gorini-Kossakowski-Sudarshan-Lindblad (GKSL) master equations for better describing the evolution of the coupled-qubit quantum heat machines~\cite{Hofer_2017, doi:10.1142/S1230161217400108}. Both of these master equations are derived in the Born-Markov approximation~\cite{PhysRevB.103.214308} and the weak-coupling approximation. For the coupled-qubit machine, we extend the local GKSL master equation to a global one whose derivation we present in Sec.~IV.A, considering the impact of coupling in the internal system. We diagonalize the system Hamiltonians of the coupled-qubit system and calculate the master equation on the transformed basis.  

On the other hand, utilizing the indirect-measurement theory \cite{PhysRevResearch.5.023066, PhysRevA.95.032132} for the whole system as we describe in Sec.~IV.~B, we quantify the work production in the work extraction process without destroying the state of the internal system. Since we use the measurement as work-production operation, we assume that the time cost in the process is negligible. 

\subsection{Global Master Equation}
Owing to the coupling between Q1 and Q2, different from the standard master equation in which each bath couples to the system in a local degree of freedom, we propose a global approach of the GKSL master equation taking the inter-dot coupling into account. 

Derivation of the global Liouville superoperators $\hat{\bar{L}}$ is more complicated than the standard one. 
To analyze the influence of the coupling to the coupled-qubit system, we diagonize the system Hamiltonian~(\ref{DQ_H}) and calculate physical quantities in the diagonalizing basis. The diagonalized system Hamiltonian $\tilde{H}_S^\alpha$ is given by 
\begin{align}
	\begin{split}
	\tilde{H}_S^\alpha&=
	U_\alpha^\dagger \cdot H_S^\alpha \cdot U_\alpha \\
	&=\begin{pmatrix}
		0 & 0 & 0 & 0 \\
		0 & \tilde{\omega}_2^\alpha & 0 & 0 \\
		0 & 0 & \tilde{\omega}_1^\alpha & 0 \\
		0 & 0 & 0 & \omega_1^\alpha+\omega_2 \\
	\end{pmatrix},
	\end{split}
\end{align}
where the eigenvalues $\tilde{\omega}_1^\alpha$ and $\tilde{\omega}_2^\alpha$ of the dressed system Hamiltonian are given by
\begin{align}
	\tilde{\omega}_1^\alpha &= \frac{1}{2}(\omega_1^\alpha+\omega_2+\sqrt{4g^2+(\omega_1^\alpha-\omega_2)^2}), \\
	\tilde{\omega}_2^\alpha &= \frac{1}{2}(\omega_1^\alpha+\omega_2-\sqrt{4g^2+(\omega_1^\alpha-\omega_2)^2}),
\end{align}
and the diagonalizing unitary transformation $U_\alpha$ is given by
\begin{align}
	U_\alpha=\begin{pmatrix}
		1 & 0 & 0 & 0 \\
		0 & \cos(\beta_\alpha) & \sin(\beta_\alpha) & 0 \\
		0 & -\sin(\beta_\alpha) & \cos(\beta_\alpha) & 0 \\
		0 & 0 & 0 & 1 \\
	\end{pmatrix}
\end{align}
with $\beta_\alpha = \theta_\alpha/2$ and $\tan(\theta_\alpha)=2g/(\omega_1^\alpha-\omega_2)$. 

The contact Hamiltonians $H_\mathrm{int}^{\alpha.q}$ between the internal coupled-qubit system and the heat baths are set to either of
\begin{align}
	H_\mathrm{int}^{\alpha,1} &= (\sigma_1^x \otimes \mathbb{I}_2) V_{k}(\hat{a}_{k, \alpha}^\dagger+\hat{a}_{k, \alpha}),\\
	H_\mathrm{int}^{\alpha,2} &= (\mathbb{I}_1 \otimes \sigma_2^x) V_{k}(\hat{a}_{k, C}^\dagger+\hat{a}_{k, C}).
\end{align}
Depending on which qubit of Q1 or Q2 contacts with the hot and cold baths, we transform the contact Hamiltonians to the diagonalizing basis as
\begin{align}
	\tilde{H}_\mathrm{int}^{\alpha,q} = U_\alpha^\dagger H_\mathrm{int}^{\alpha,q} U_\alpha, 
\end{align}
where $q=1,2$ indicates the qubit of the internal system. 

We conduct all the calculations in the diagonalizing basis, so that the transformed interactions between the qubit and the heat baths are given by
\begin{align}
	(\tilde{\sigma}^\alpha)_1^x(t) =&e^{i\tilde{H}_S^\alpha t} \tilde{H}_\mathrm{int}^{\alpha,1} e^{-i\tilde{H}_S^\alpha t} \\
	\begin{split}
	=& \cos(\beta)\tilde{\mathbb{I}}^\alpha_2((\tilde{\sigma}^\alpha)_1^+e^{-i\tilde{\omega}_1^\alpha t} + (\tilde{\sigma}^\alpha)_1^- e^{i\tilde{\omega}_1^\alpha t})\\
	&-\sin(\beta)(\tilde{\sigma}^\alpha)_1^z((\tilde{\sigma}^\alpha)_2^+ e^{-i\tilde{\omega}_2^\alpha t} + (\tilde{\sigma}^\alpha)_2^- e^{i\tilde{\omega}_2^\alpha t}),\\
	\end{split}\\
	(\tilde{\sigma}^\alpha)_2^x(t) =&e^{i\tilde{H}_S^\alpha t} \tilde{H}_\mathrm{int}^{\alpha,2} e^{-i\tilde{H}_S^\alpha t} \\
	\begin{split}
	=& \sin(\beta)(\tilde{\sigma}^\alpha)_2^z((\tilde{\sigma}^\alpha)_1^+e^{-i\tilde{\omega}_1^\alpha t} +(\tilde{\sigma}^\alpha)_1^- e^{i\tilde{\omega}_1^\alpha t})\\
	&+\cos(\beta)\tilde{\mathbb{I}}^\alpha_1((\tilde{\sigma}^\alpha)_2^+ e^{-i\tilde{\omega}_2^\alpha t} + (\tilde{\sigma}^\alpha)_2^- e^{i\tilde{\omega}_2^\alpha t}),
	\end{split}
\end{align}
where $(\tilde{\sigma}^\alpha)_q^k$ and $\tilde{\mathbb{I}}^\alpha_q$ are the $k$ ($k=x,y,z,+,-$) component of the Pauli matrices and the identity matrix in the diagonizing basis, repectively. The global master equation
\begin{align}
	&\frac{d\rho}{dt}=-i[H,\rho]+\hat{\bar{L}}_\alpha\rho,
\end{align}
and the Liouville superoperators are transformed as 
\begin{align}
\begin{split}
	\hat{\tilde{L}}_{\alpha,1}\rho =&[\cos[2](\beta^\alpha)G_H(\tilde{\omega}_1^\alpha)\hat{D}[\tilde{\sigma^\alpha}_1^-] \\
	&+\cos[2](\beta^\alpha)G_\alpha(-\tilde{\omega}_1^\alpha)\hat{D}[\tilde{\sigma^\alpha}_1^+] \\
	&+\sin[2](\beta^\alpha)G_\alpha(\tilde{\omega}_2^\alpha)\hat{D}[\tilde{\sigma^\alpha}_2^-] \\
	&+\sin[2](\beta^\alpha)G_\alpha(-\tilde{\omega}_2^\alpha)\hat{D}[\tilde{\sigma^\alpha}_2^+]]\rho,
\end{split}\\
\begin{split}
	\hat{\tilde{L}}_{\alpha,2}\rho =&[\sin[2](\beta^\alpha)G_\alpha(\tilde{\omega}_1^\alpha)\hat{D}[\tilde{\sigma^\alpha}_1^-] \\
	&+\sin[2](\beta^\alpha)G_\alpha(-\tilde{\omega}_1^\alpha)\hat{D}[\tilde{\sigma^\alpha}_1^+] \\
	&+\cos[2](\beta^\alpha)G_\alpha(\tilde{\omega}_2^\alpha)\hat{D}[\tilde{\sigma^\alpha}_2^-] \\
	&+\cos[2](\beta^\alpha)G_\alpha(-\tilde{\omega}_2^\alpha)\hat{D}[\tilde{\sigma^\alpha}_2^+]]\rho.
\end{split}
\end{align}
Note that the definitions of  the Lindblad dissipators and other quantities are the same as in the single-qubit case given by Eq.~(\ref{Lindblad})--(\ref{SDF}). 

With the assistant of Python quantum tool Qutip \cite{qutip1,qutip2}, we simulate the models and compare them from several angles. We follow the standard definitions for heat flowing, obtaining 
\begin{align}
	&Q_h = \tr[\tilde{H}_S^h(\rho(t_h)-\rho(0))], \\
	&Q_c = \tr[\tilde{H}_S^c(\rho(t_c+t_h)-\rho(t_h))],
\end{align}
where $t_\alpha$ denote the time costs of the interaction between the internal coupled-qubit system and each heat bath $\alpha$. The definition of the work production is given below.

\subsection{Work Extraction Process}
Extending the indirect-measurement method in Refs.~\cite{PhysRevResearch.5.023066, PhysRevA.95.032132} for the calculation of work production in the single-qubit heat engine, we come up with a method for the coupled-qubit heat engine in our scheme. We update the work-production process in Sec.~II.~C by transforming it to the diagonalizing basis and changing the work storages from the single two-level system for the single-qubit cycle to the double two-level systems for our coupled-qubit system. 
We also let $t_\alpha$ denote the time costs of the interaction between the internal coupled-qubit system and each heat bath $\alpha$. We will obtain the work production that still satisfies the standard definition:
\begin{align}
	\label{W1}
	&W_1 = \tr[\rho(t_h)(\tilde{H}_S^c-\tilde{H}_S^h)], \\
	\label{W2}
	&W_2 = \tr[\rho(t_h+t_c)(\tilde{H}_S^h-\tilde{H}_S^c)].
\end{align}
Since the work extraction process is achieved by the indirect measurement, we assume that it takes negligible time.

For its derivation, we introduce a one-qubit clock and a two-qubit work storage so that we can observe the work production of the internal system through the measurement of the work storage but do not destroy the system state. To consider the influence of the coupling, which plays a key role in our research, we conduct the measurement in the transformed basis with the diagonalized Hamiltonian
\begin{align}
\begin{split}
	\tilde{H}_S^\alpha = &\tilde{\omega}_2^\alpha \ket{\downarrow\uparrow}_S\bra{\downarrow\uparrow} + \tilde{\omega}_1^\alpha \ket{\uparrow\downarrow}_S\bra{\uparrow\downarrow} \\&+ (\omega_1^\alpha+\omega_2) \ket{\uparrow\uparrow}_S\bra{\uparrow\uparrow}.
\end{split}
\end{align}
In the instance of the step~(c) in Fig.~\ref{cycles} that the system Hamiltonian changes from $\bar{H}_S^h$ to $\bar{H}_S^c$, the entire Hamiltonian of the internal system and external environment after the introduction of the clock and the work storage is given by
\begin{align}
\begin{split}
	H_{SE} = &\tilde{H}_S^h  \otimes \ket{0}_C\bra{0} \otimes \mathbb{I}_W \\&+\tilde{H}_S^c\otimes\ket{1}_C\bra{1}\otimes\mathbb{I}_W +H_W, 
\end{split}
\end{align}
where $H_W$ is the Hamiltonian of the work storage given by
\begin{align}
\begin{split}
	H_W &= \mathbb{I}_S \otimes \mathbb{I}_C \otimes
	\big[ (\tilde{\omega}_2^C-\tilde{\omega}_2^H)\ket{\downarrow\uparrow}_W\bra{\downarrow\uparrow} \\&+ (\tilde{\omega}_1^C-\tilde{\omega}_1^H)\ket{\uparrow\downarrow}_W\bra{\uparrow\downarrow} + (\omega_1^C-\omega_1^H)\ket{\uparrow\uparrow}_W\bra{\uparrow\uparrow} \big].
\end{split}
\end{align}
The unitary transformation is given by
\begin{align}
\begin{split}
	U_W = &\ket{\downarrow\downarrow}_S\bra{\downarrow\downarrow}\otimes (\ket{0}_C\bra{1}+\ket{1}_C\bra{0})  \otimes \mathbb{I}_W \\
	&+\sum_{b=\downarrow\uparrow,\uparrow\downarrow,\uparrow\uparrow} \ket{b}_S\bra{b} \otimes \big[\ket{0}_C\bra{1} \otimes \ket{\downarrow\downarrow}_W\bra{b}\\
	& + \ket{1}_C\bra{0} \otimes \ket{b}_W\bra{\downarrow\downarrow}\\
	&+\ket{0}_C\bra{0} \otimes (\mathbb{I}_W-\ket{\downarrow\downarrow}_W\bra{\downarrow\downarrow}) \\
	&+ \ket{1}_C\bra{1} \otimes (\mathbb{I}_W-\ket{b}_W\bra{b})\big],
\end{split}
\end{align}
where $\ket{b}_S\bra{b}$, $(b=\downarrow\downarrow,\downarrow\uparrow,\uparrow\downarrow,\uparrow\uparrow)$ are the eigenbases of our coupled-qubit system.
Since $U_W$ commutes with the entire Hamiltonian $H_{SE}$, the energy in this process of work extraction is conserved, which satisfies the first thermodynamical law. The initial and final density matrices of the entire state are given by
\begin{align}
	\rho_i &= \sum_{b}p_{b}\ket{b}_S\bra{b} \otimes \ket{0}_C\bra{0} \otimes \ket{\downarrow\downarrow}_W\bra{\downarrow\downarrow},\\
	\rho_f &= \sum_{b}p_{b}\ket{b}_S\bra{b} \otimes \ket{1}_C\bra{1} \otimes \ket{w}_W\bra{w},
\end{align}
respectively, where $p_{b}$ is the probability for the internal system existing in each eigenstate. 

We do the projection measurement on the work storage and find its state as 
\begin{align}
	\rho_W = \sum_{w}p_{w}\ket{w}_W\bra{w},
\end{align}
which gives the energy of the work storage in the form
\begin{align}
	W =& \tr[H_W \rho_W] \\
	\begin{split}
	    =& p_{\downarrow\uparrow}(\tilde{\omega}_2^H-\tilde{\omega}_2^C)+p_{\uparrow\downarrow}(\tilde{\omega}_1^H-\tilde{\omega}_1^C)+p_{\uparrow\uparrow}(\omega_1^H-\omega_1^C)
	    \end{split}\\
	    = &\tr[\rho(t_h)(\tilde{H}_S^c-\tilde{H}_S^h)] = W_1.
\end{align}
This indeed is equivalent to Eq.~(\ref{W1}) based on the standard definition. We can similarly derive Eq.~(\ref{W2}), employing the indirect-measurement scheme.
The power and the efficiency of the engine are defined in Eqs.~(\ref{P}) and (\ref{Eff}).

\section{Numerical Calculation}
Hitherto, properties of the single-qubit Otto engine are summarized in Sec.~II, and the four schemes of our coupled-qubit Otto engine and their main dynamics are explained in Secs.~III and IV. 
In this section, we numerically analyze the coupled-qubit system and compare its maximum power to that of the single-qubit engine. In Secs.~IV.~A--D, we analyze the results of the four models of our coupled-qubit engine and find the maximum power. We mainly focus on Model~12 and Model~21 in Secs.~IV.~A and~B, respectively, which are the most interesting parts among our coupled-qubit models. We also analyze Model~11 and Model~22 in Secs.~IV.~C and~D, respectively, which might be also useful in some applications. In Sec.~IV.~E, we compare the four coupled-qubit engines to the single-qubit engine, which demonstrates the effects of the coupling on the Otto engine, and make the comparison of the coupled-qubit systems in different situations, which plays a key role for versatile applications. 

For our numerical simulations of the coupled-qubit Otto cycle, we set relevant parameters by assuming the energy unit as $E_{\mathrm{unit}}=\omega_c=\omega_2=1$. We also fix the transition rate as $\kappa_h=\kappa_c=0.005$ and the time costs of the evolution in the ischoric processes as $t_c=t_h=50$, exactly the same as in the analysis of the single-qubit case. Similarly to the maximum power of the single-qubit engine, which happens as the peak of the power depending on the energy levels $\omega_\alpha$ of the internal system under specific temperatures $T_\alpha$ of the heat baths, we define the maximum power of our coupled-qubit engine as the peak of the power depending on the energy level $\omega_1^c$ of Q1 under specific temperatures $T_\alpha$ of the heat baths and the coupling strength $g$. Note that as indicated in Fig.~\ref{DQ_E}, we fix $\omega_1^h$ according to Eq.~(\ref{MPR2}).

\subsection{Model~12}
As defined in Sec.~III.~A, Q1 and Q2 of the Model~12 interacts with the hot and the cold bath, respectively, in the ischoric processes. In the processes of the work production, on the other hand, the energy level $\omega_2$ of Q2 is fixed and the work storages interact with the internal system only through Q1.
\begin{figure*}
	\centering
	\subcaptionbox[subcaption1]{Power\label{M12:power}}[0.49\textwidth]{
		\includegraphics[scale=0.75]{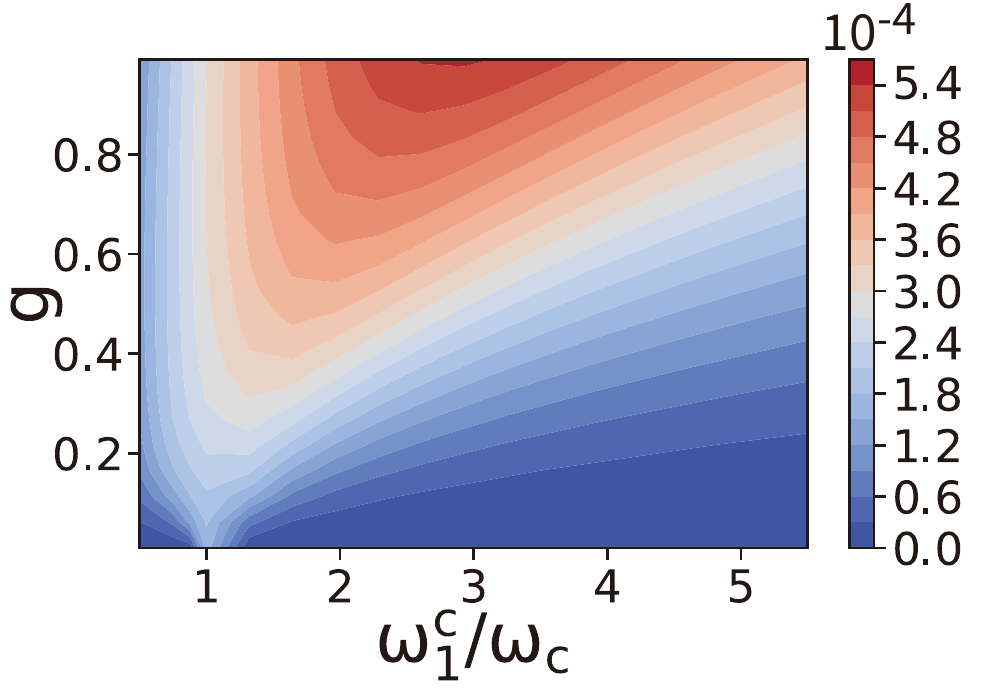}}\hfill
	\subcaptionbox[subcaption3]{Efficiency\label{M12:efficiency}}[0.49\textwidth]{
		\includegraphics[scale=0.75]{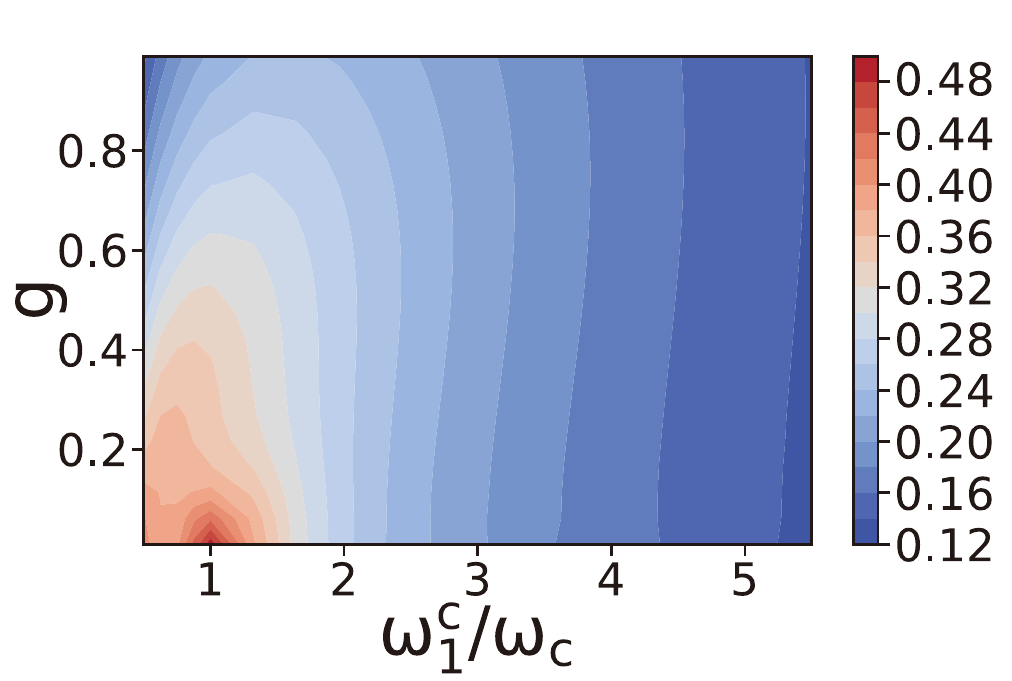}}
	\caption{Dependence of (a)~the power $P$ and (b)~the efficiency $\eta$ of Model~12 on the Q1's energy level $\omega_1^c/\omega_c$ and the coupling strength $g$ under the fixed heat-bath temperatures. Parameters: energy unit $\omega_c=1$; temperature of heat baths $T_c = 5, T_h=15$; transition rate: $\kappa_h = \kappa_c = 0.005$; time durations: $t_h = t_c = 50$.}
	\label{M12_density}
\end{figure*}

\begin{figure*}
	\centering
	\subcaptionbox[subcaption1]{\label{M12:P_kw}}[0.49\textwidth]{
		\includegraphics[scale=0.9]{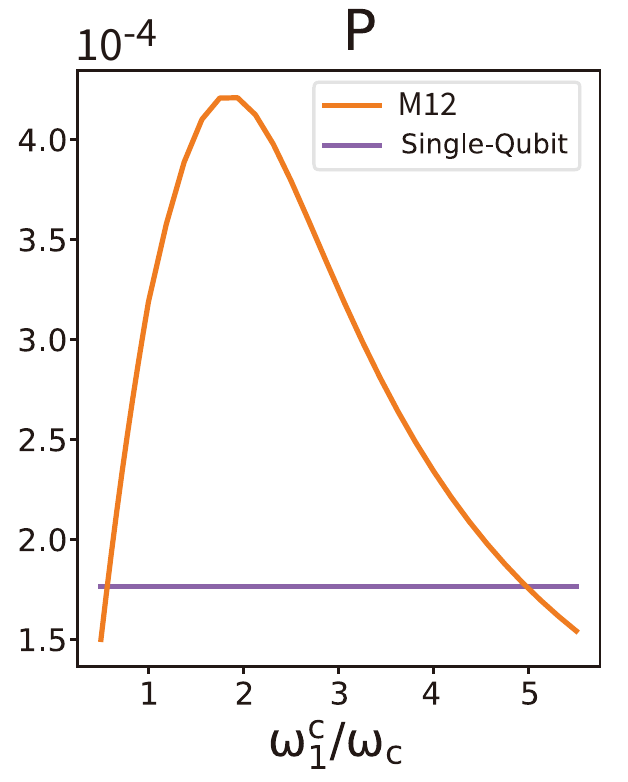}}\hfill
	\subcaptionbox[subcaption2]{\label{M12:Eff_kw}}[0.49\textwidth]{
		\includegraphics[scale=0.9]{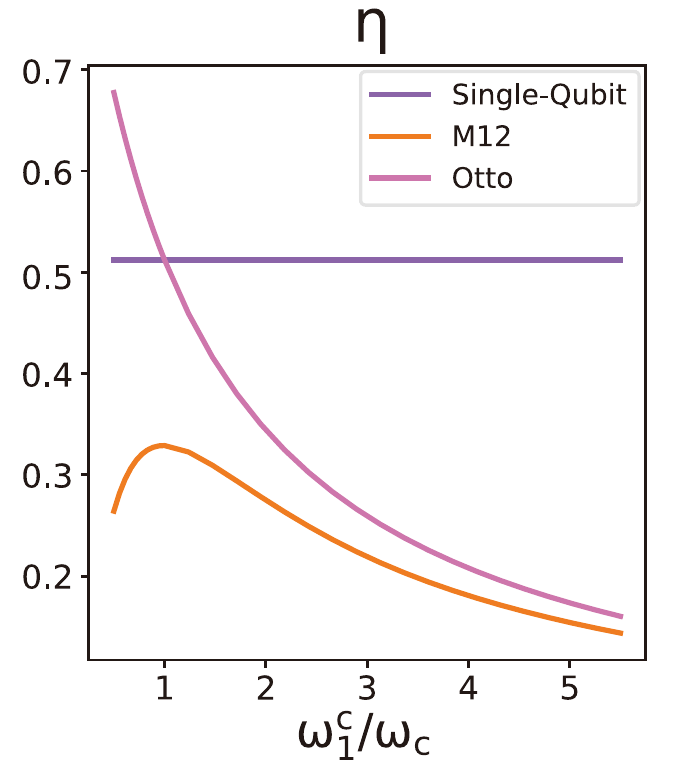}}\par
\caption{Dependence of (a)~the power $P$ and (b)~efficiency $\eta$ of Model~12 on the Q1's energy level $\omega_1^c/\omega_c$  for the fixed coupling strength $g=0.55$. The blue lines indicates the results of the single-qubit engine for comparisonParameters: energy unit $\omega_c = 1$; temperature of heat baths $T_c = 5, T_h=15.5$; transition rate $\kappa_h = \kappa_c = 0.005$; time durations $t_h = t_c = 50$.}
	\label{M12_kwg}
\end{figure*}

As shown in Fig.~\ref{M12_density}, for the fixed energy levels of Q1, the power of Model~12 increases and approaches to a greatest value when the coupling strength $g$ gets stronger. On the other hand, for a fixed coupling strength $g$, the power of Model~21 increases first but decreases after a peak when the energy level $\omega_1^c$ of Q1 increases, and we can find a peak of power in Model~12 depending on the energy levels of Q1, as shown in Fig.~\ref{M12:P_kw}, which we define as the maximum power of Model~12 for specific coupling strength $g=0.55$ and heat-bath temperatures $T_c=5$, $T_h = 15.5$; we will hereafter use the values for comparison. As shown in Fig.~\ref{M12:Eff_kw}, the efficiency of Model~12 is lower than its Otto efficiency while the coupling improves the power, unlike the single-qubit system, for which the efficiency at the maximum power is equal to its Otto efficiency; see Eq.~(\ref{Otto})

\begin{figure*}
	\flushleft
	\subcaptionbox[subcaption1]{$P_m$\label{M12:Pm}}[0.27\textwidth]{
		\includegraphics[scale=0.75]{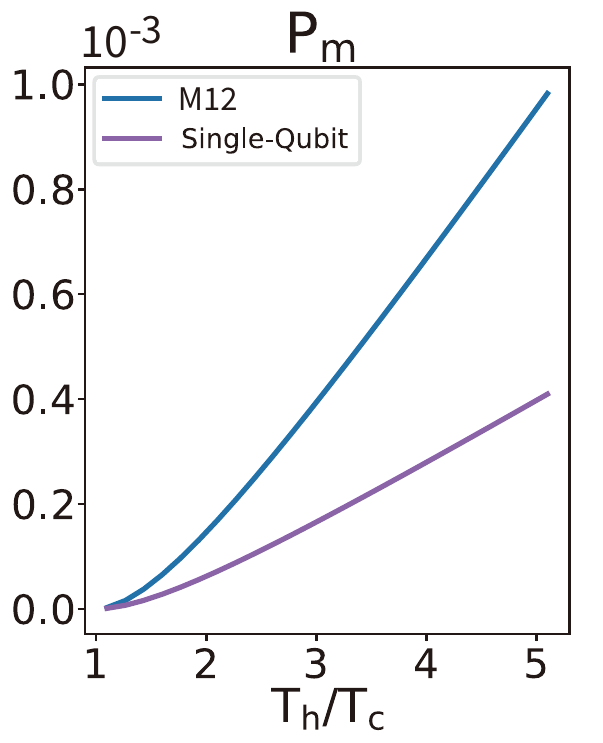}}
	\subcaptionbox[subcaption3]{$(\omega_1^c/\omega_c)_{P_m}$\label{M12:kw}}[0.28\textwidth]{
		\includegraphics[scale=0.75]{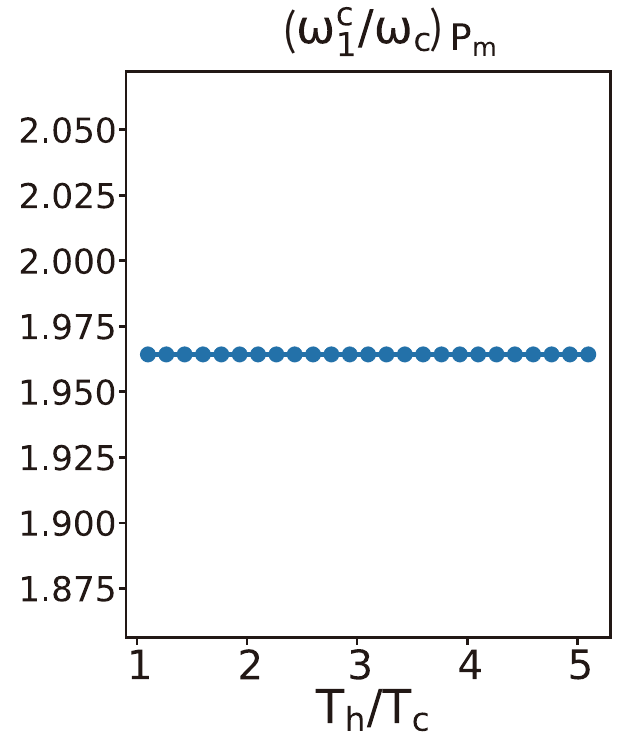}}
	\subcaptionbox[subcaption4]{$\eta_{P_m}$\label{M12:Eff}}[0.33\textwidth]{
		\includegraphics[scale=0.75]{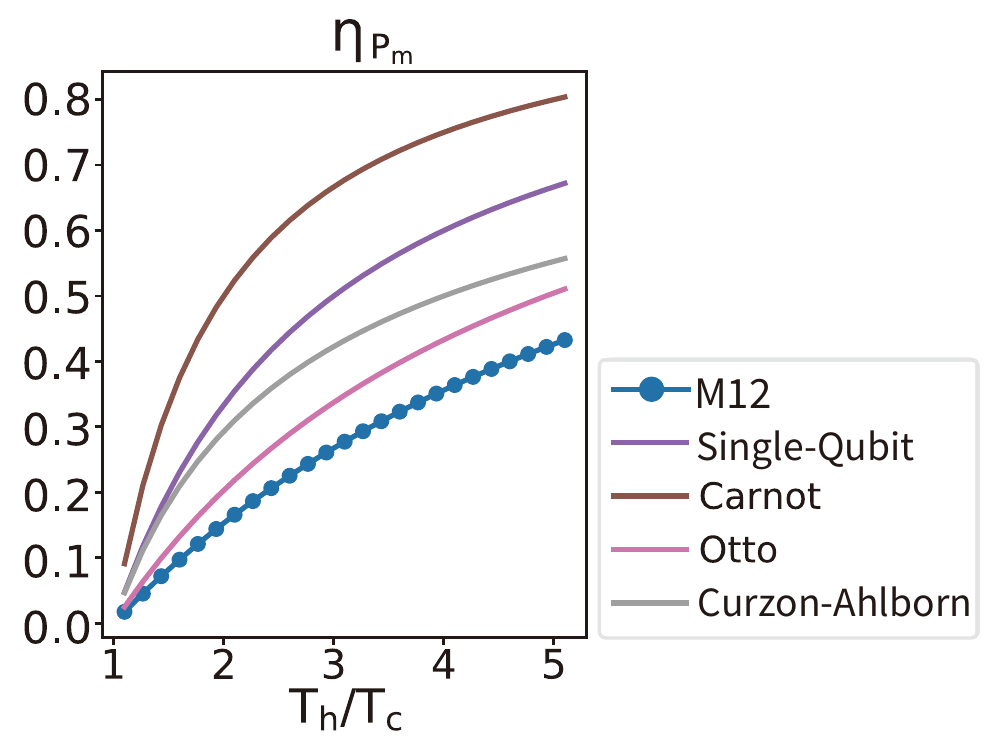}}
	\caption{Dependence of (a)~the maximum power $P_m$, (c)~the energy level $(\omega_1^c/\omega_c)_{P_m}$ of Q1 and (d)~the efficiency $\eta_{P_m}$ of Model~12 on the heat-bath temperatures $T_h/T_c$ for the fixed coupling strength $g=0.55$. Parameters: energy unit $\omega_c=1$; temperature of cold bath $T_c = 5$; transition rate $\kappa_h = \kappa_c = 0.005$; time cost $t_h = t_c = 50$.}
	\label{M12_Pm}
\end{figure*}

Figure~\ref{M12:Pm} presents the maximum power of the Model~12 for different temperatures of the heat baths $T_\alpha$ when the coupling strength $g$ is fixed to 0.55. The energy level $\omega_1^c$ of Q1 for achieving the maximum power remains constant for different heat-bath temperatures, as shown in Fig.~\ref{M12:kw}. Although the level change $\Delta \omega$ of our coupled-qubit models is set to be equal to the one that maximizes the power of the single-qubit system as in Eq.~(\ref{MPR3}), Model~12 achieves much greater powers when Q1's energy level $\omega_1^c$ is about two times higher than the energy levels of the single-qubit system $\omega_c$. In other words, Model~12 breaks the maximum-power relation~(\ref{MPR1}) of the single-qubit engine and achieves a greater maximum power with higher energy levels, thanks to the existence of the other qubit Q2 and the coupling between the two qubits.

Comparison of different efficiencies is shown in Fig.~\ref{M12:Eff}. As mentioned in Sec.~II.~D, the efficiency of the single-qubit system is equal to its Otto efficiency (purple line) at the maximum power as Eq.~(\ref{Otto}). Since the energy level $\omega_1^c$ of Q1 is higher than the energy level $\omega_c$ of the single-qubit system, the Otto efficiency of Model~12 (pink line) is lower than the single-qubit Otto efficiency (purple line). Besides, as mentioned before, when the coupling improves the power of the coupled-qubit engine, the efficiency at the maximum power decreases, which is the reason why the system efficiency of Model~12 (blue line) is lower than the Otto efficiency of Model~12 (pink line). In other words, similar to the Curzon-Ahlborn efficiency (grey line), which is the efficiency at the maximum power of the Carnot cycle and lower than the Carnot efficiency (brown line), the efficiency at the maximum power of Model~12 is lower than its Otto efficiency. As a result, the system efficiency of Model~12 (blue line) at the maximum power for a specific coupling strength $g=0.55$ and heat-bath temperatures $T_\alpha$ is lowest among the efficiencies listed above.   
 
\subsection{Model~21}
As defined in Sec.~III.~A, for Model~21, Q2 and Q1 interact with the hot and the cold baths, respectively, in the ischoric processes. In the processes of work production, the energy level $\omega_2$ of Q2 is fixed and the work storages interact with the internal system only through Q1.

\begin{figure*}
	\centering
	\subcaptionbox[subcaption1]{Power\label{M21:power}}[0.49\textwidth]{
		\includegraphics[scale=0.75]{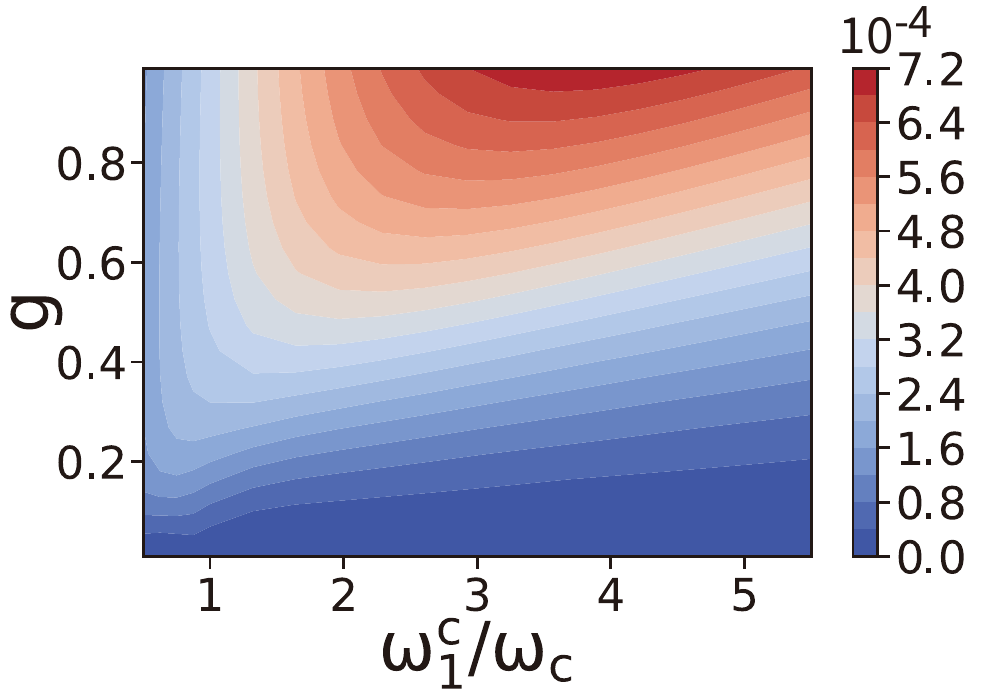}}
	\subcaptionbox[subcaption3]{Efficiency\label{M21:efficiency}}[0.49\textwidth]{
		\includegraphics[scale=0.75]{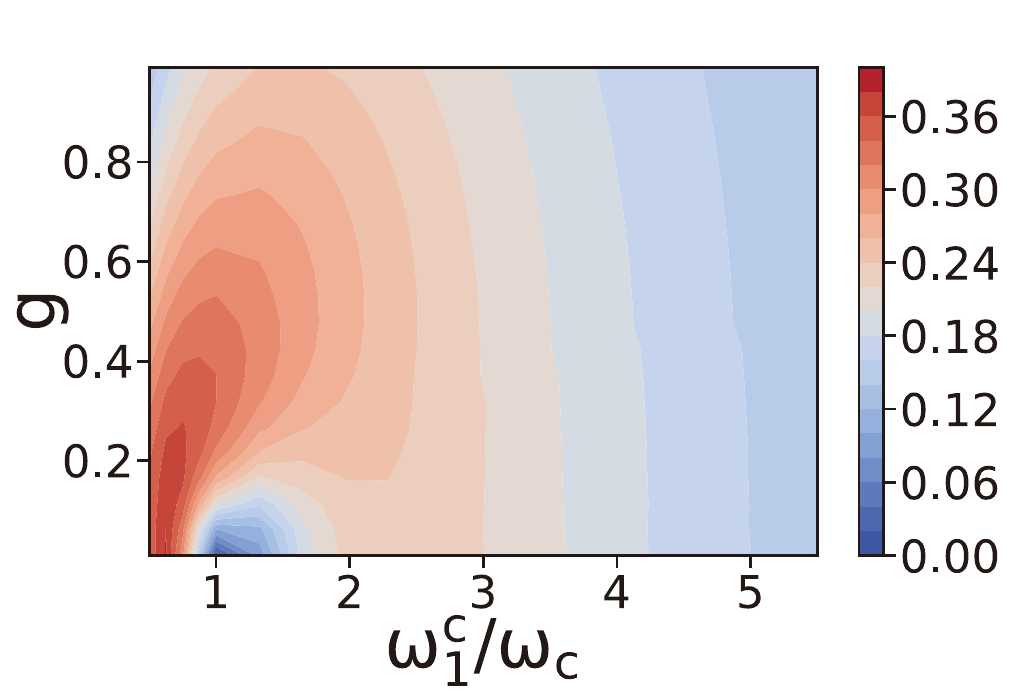}}
	\caption{Dependence of (a)~the power $P$ and (b)~the efficiency $\eta$ of Model~21 on the Q1's energy level $\omega_1^c/\omega_c$ and the coupling strength $g$ under the fixed heat-bath temperatures. Parameters: energy unit $\omega_c=1$; temperature of heat baths $T_c = 5, T_h=15$; transition rates: $\kappa_h = \kappa_c = 0.005$; time durations: $t_h = t_c = 50$}
	\label{M21_density}
\end{figure*}

\begin{figure*}
	\centering
	\subcaptionbox[subcaption1]{\label{M21:P_kw}}[0.49\textwidth]{
		\includegraphics[scale=0.8]{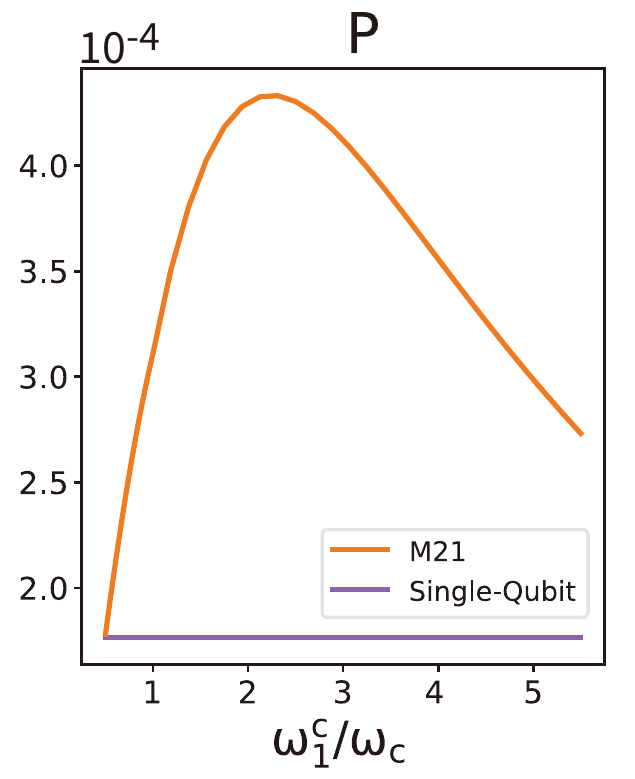}}\hfill
	\subcaptionbox[subcaption2]{\label{M21:Eff_kw}}[0.49\textwidth]{
		\includegraphics[scale=0.8]{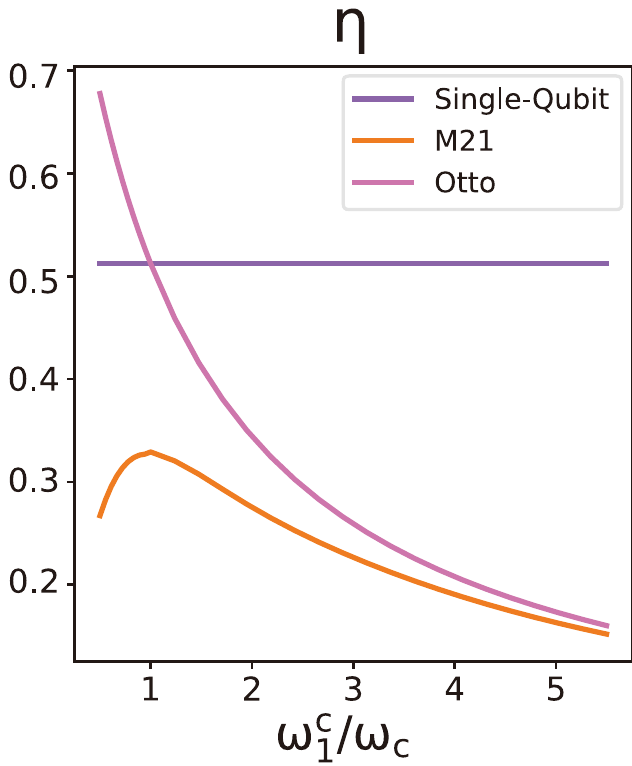}}
	\caption{Dependence of (a)~the power $P$ and (b)~efficiency $\eta$ of Model~21 on the Q1's energy level $\omega_1^c/\omega_c$ for the fixed coupling strength $g=0.55$. The blue lines indicates the results of the single-qubit engine for comparison. Parameters: energy unit $\omega_c = 1$; temperature of heat baths $T_c = 5, T_h=15.5$; transition rates $\kappa_h = \kappa_c = 0.005$; time durations $t_h = t_c = 50$}
	\label{M21_kwg}
\end{figure*}

Similarly to the case of Model~12, we search for the maximum power of Model~21 by adjusting the energy levels $\omega_1^\alpha$ of Q1 and the coupling strength $g$ for fixed temperature $T_\alpha$ of the heat baths and the energy-level change $\Delta\omega$; see Fig.~\ref{M21_density}. The dependence of the power of Model~21 on the coupling strength $g$ and the energy level $\omega_1^c$ of Q1 is similar to the case of Model~12. For the fixed energy levels of Q1, the power of Model~21 increases when the coupling strength gets stronger. On the other hand, for a fixed coupling strength, the power of Model~21 increases first but decreases after a peak when the energy levels $\omega_1^\alpha$ of Q1 increase.
Therefore, as shown in Fig.~\ref{M21:P_kw}, we can find a peak of power depending on the energy levels of Q1 by fixing the coupling strength for specific temperatures of the heat baths, which we define as the maximum power of Model~21 for specific coupling strength $g$ and heat-bath temperatures $T_\alpha$. Similarly to Model~12, as shown in Fig.~\ref{M21:Eff_kw}, the system efficiency of Model~21 is lower than the one at the maximum power of the single-qubit system for the specific heat-bath temperatures and the Otto efficiency of Model~21 for various values of the energy level $\omega_1^c$ of Q1. The coupling decreases the efficiency when it improves the power of Model~21.

\begin{figure*}
	\flushleft
	\subcaptionbox[subcaption1]{$P_m$\label{M21:Pm}}[0.27\textwidth]{
		\includegraphics[scale=0.77]{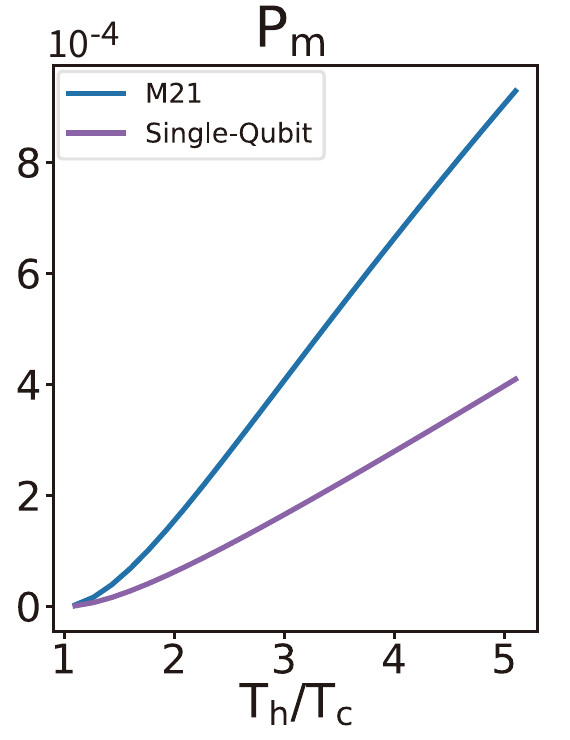}}
	\subcaptionbox[subcaption3]{$(\omega_1^c/\omega_c)_{P_m}$\label{M21:kw}}[0.28\textwidth]{
		\includegraphics[scale=0.77]{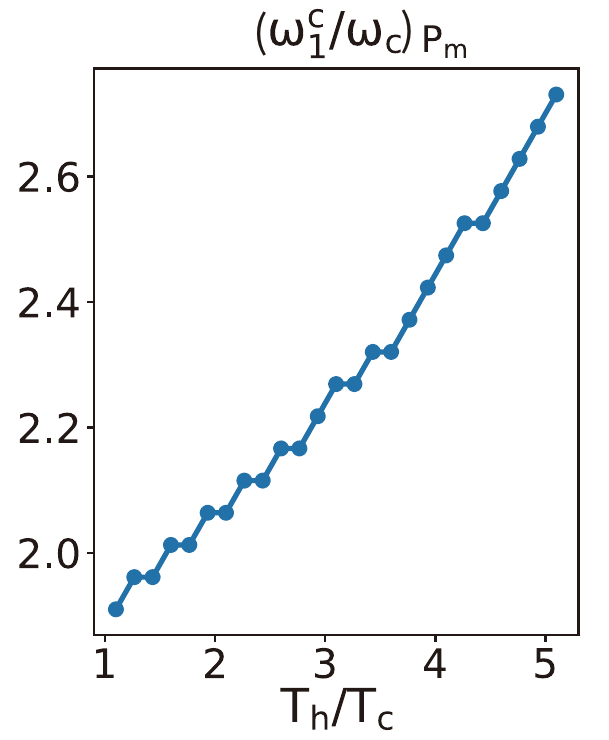}}
	\subcaptionbox[subcaption4]{$\eta_{P_m}$\label{M21:Eff}}[0.33\textwidth]{
		\includegraphics[scale=0.77]{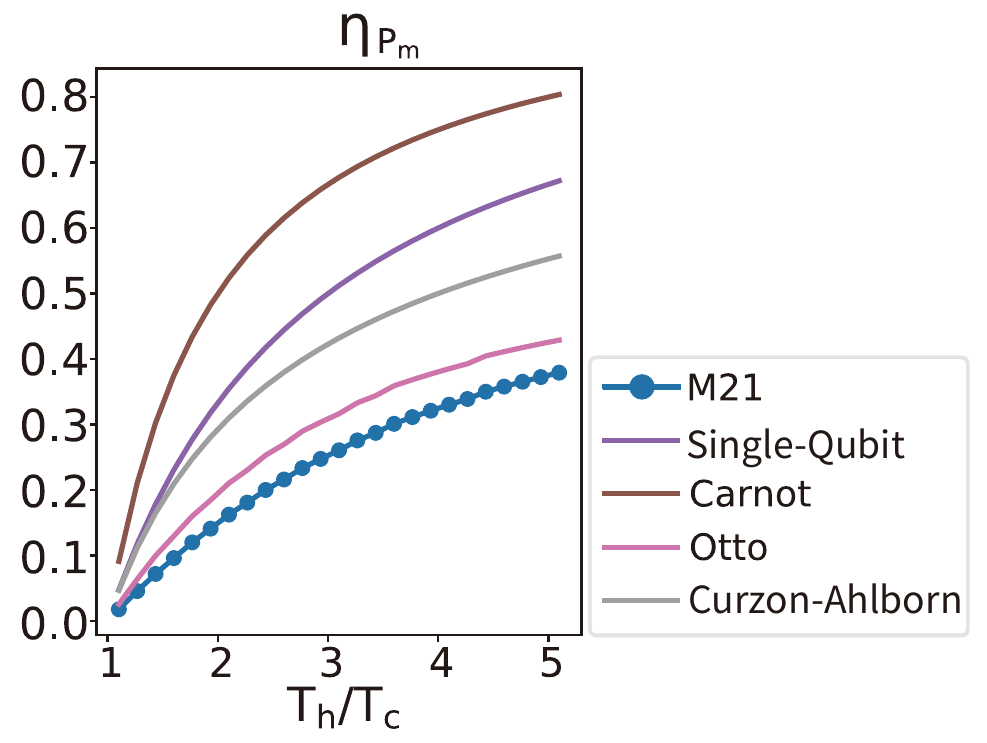}}
	\caption{Dependence of (a)~the maximum power $P_m$, (b)~the energy level $(\omega_1^c/\omega_c)_{P_m}$ of Q1 and (c)~the efficiency $\eta_{P_m}$ of Model~21 on the heat-bath temperatures $T_h/T_c$ for the fixed coupling strength $g=0.55$. Parameters: energy unit $\omega_c=1$; temperature of cold bath $T_c = 5$; transition rates: $\kappa_h = \kappa_c = 0.005$; time durations: $t_h = t_c = 50$}
	\label{M21_Pm}
\end{figure*}

Similarly to the case of Model~12 in Fig.~\ref{M12:Pm}, although the level change $\Delta \omega$ of Model~21 is equal to the one which maximizes the power of the single-qubit system, Model~21 also achieves much greater powers than the single-qubit one. However, unlike the Model~12, for which the energy levels of Q1 for achieving the maximum power remain constant (Fig.~\ref{M12:kw}) for different heat-bath temperatures, for Model~21 in Fig.~\ref{M21:kw}, the energy level $\omega_1^c$ of Q1 achieving the maximum power becomes higher when the ratio $T_h/T_c$ of the heat-bath temperatures increases. Model~21 also breaks the maximum-power relation~(\ref{MPR1}) of the single-qubit system and achieves much greater powers than the single-qubit one when its Q1's energy level $\omega_1^c$ is higher than the energy level of the single-qubit system $\omega^c$.

For the comparison of different efficiencies in Fig.~\ref{M21:Eff}, the efficiency of Model~21 plotted by the blue line is again the lowest among the efficiency listed here. Due to the influence of the $XX$-coupling on the coupled-qubit system, the system efficiency of Model~21 at the maximum power (blue line) is lower than its Otto efficiency (pink line), which is similar to the case of the Carnot cycle in that the Curzon-Ahlborn efficiency (grey line) is lower than the Carnot efficiency (brown line). Besides, the Otto efficiency of Model~21 at the maximum power (blue line) is lower than the single-qubit system's Otto efficiency (purple line), due to the higher energy level $\omega_1^c$ of Q1 than the energy level $\omega_c$ of the single-qubit system. Unlike the single-qubit system, for which the system efficiency at the maximum power is higher than the Curzon-Ahlborn efficiency, the system efficiency of Model~21 at the maximum power is lower than the Curzon-Ahlborn efficiency, as the blue and purple lines show in Fig.~\ref{M21:Eff}.

\subsection{Model~11}
\begin{figure*}
	\centering
	\subcaptionbox[subcaption1]{Power\label{M11:power}}[0.49\textwidth]{
		\includegraphics[scale=0.75]{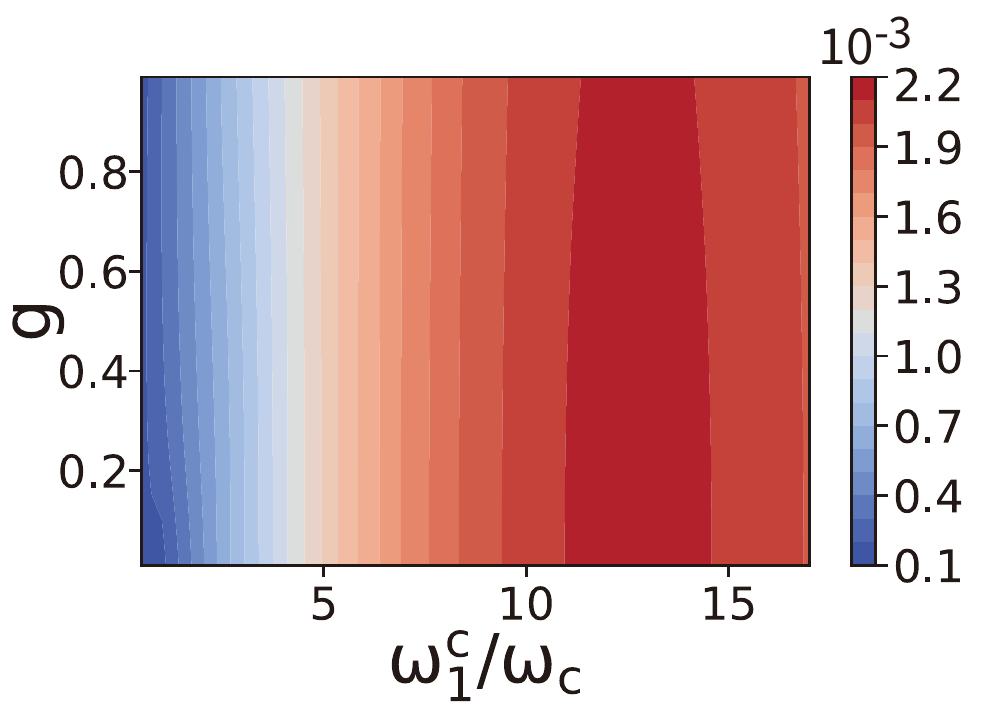}}\hfill
	\subcaptionbox[subcaption3]{Efficiency\label{M11:efficiency}}[0.49\textwidth]{
		\includegraphics[scale=0.75]{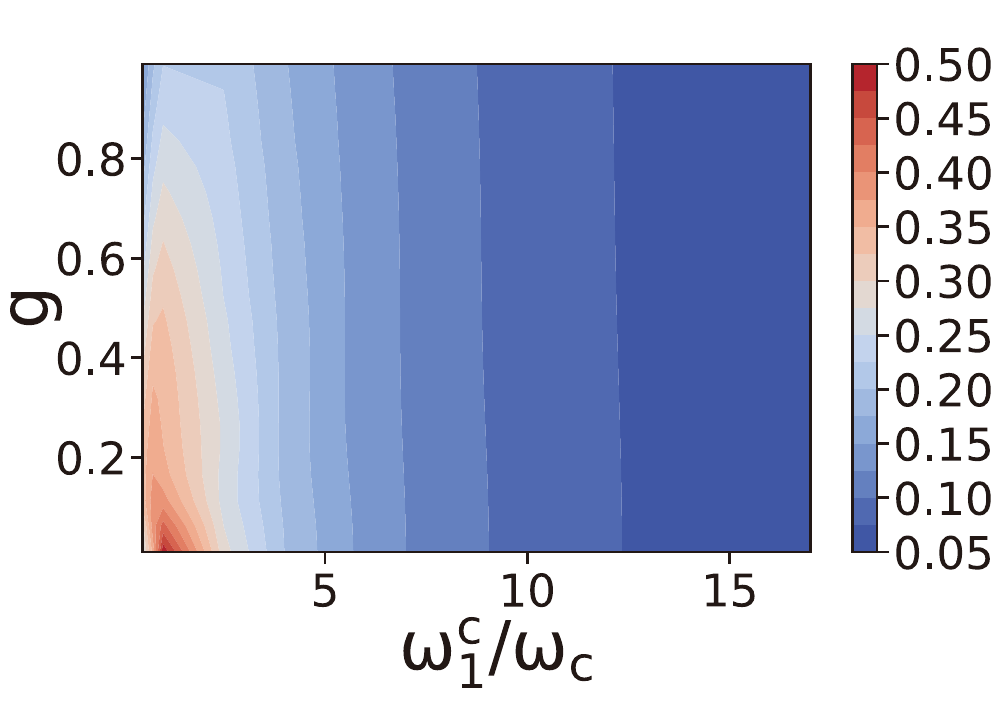}}
	\caption{Dependence of (a)~the power $P$ and (b)~the efficiency $\eta$ of Model~11 on the Q1's energy level $\omega_1^c/\omega_c$ and the coupling strength $g$ under the fixed heat-bath temperatures. PARAMETERS: transition rate: $\kappa_h = \kappa_c = 0.005$, time cost: $t_h = t_c = 50$, Energy unit $\omega_c=1$, temperature of heat baths $T_c = 5, T_h=15$}
	\label{M11_density}
\end{figure*}

\begin{figure*}
	\centering
	\subcaptionbox[subcaption1]{power\label{M11:P_kw}}[0.49\textwidth]{
		\includegraphics[scale=0.9]{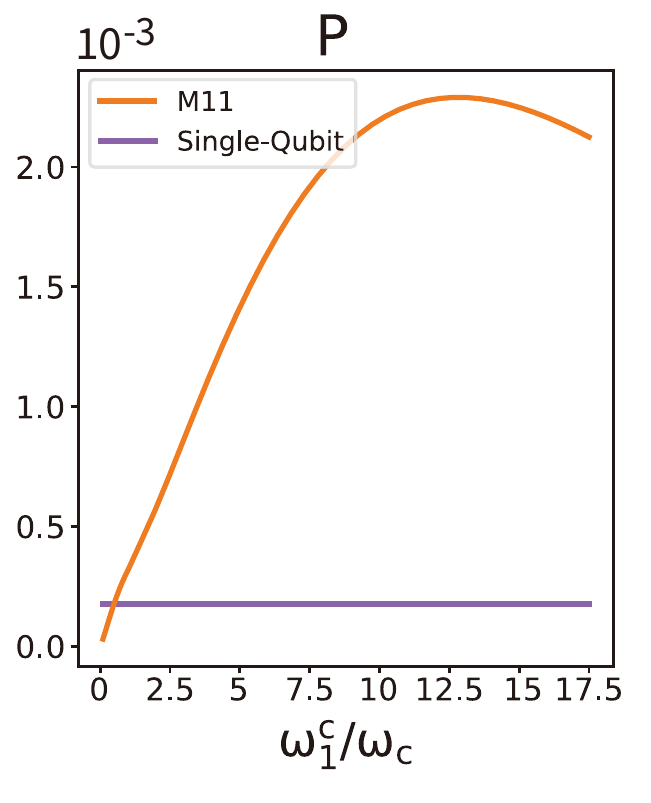}}\hfill
	\subcaptionbox[subcaption2]{efficiency\label{M11:Eff_kw}}[0.49\textwidth]{
		\includegraphics[scale=0.9]{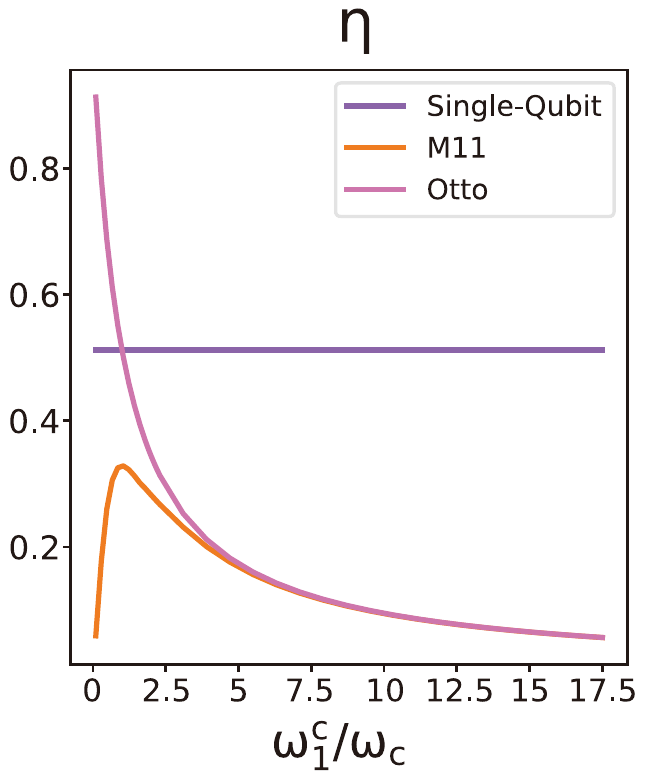}}\par
	\caption{Dependence of (a)~the power $P$ and (b)~efficiency $\eta$ of Model~11 on the Q1's energy level $\omega_1^c/\omega_c$ for the fixed coupling strength $g=0.55$. The blue line indicates the results of the single-qubit engine for comparison. Parameters: energy unit: $\omega_c = 1$; temperature of heat baths: $T_c = 5, T_h=15.5$; transition rates: $\kappa_h = \kappa_c = 0.005$; time durations: $t_h = t_c = 50$}
	\label{M11_kwg}
\end{figure*}

As defined in Sec.~III.~A, Model~11 interacts with each heat bath only through Q1 during the ischoric processes and with the work storages only through Q1 in the processes of work production. 
We again search for the maximum power of Model~11 by adjusting the energy levels $\omega_1^\alpha$ of Q1 and the coupling strength $g$ for fixed temperatures $T_\alpha$ of heat baths and the energy-level change $\Delta\omega$ of Q1 in Eq.~(\ref{MPR3}); see Fig.~\ref{M11_density}. The dependence of the power on the energy levels of Q1 with fixed coupling strength is similar to the previous models in that the power increases first but decreases after a peak when the energy levels of Q1 increase with a specific coupling strength, as shown in Fig.~\ref{M11:P_kw}. Therefore, by fixing the coupling strength to a constant value, we can still define the peak of the power depending on the energy levels of Q1 as the maximum power of Model~11 under specific heat-bath temperatures and coupling strength. However, if we fix the energy level of Q1 to the one which maximizes the power, unlike Model~12 and Model~21, the power and efficiency are almost independent of the coupling strength $g$.

\begin{figure*}
	\flushleft
	\subcaptionbox[subcaption1]{$P_m$\label{M11:Pm}}[0.27\textwidth]{
		\includegraphics[scale=0.77]{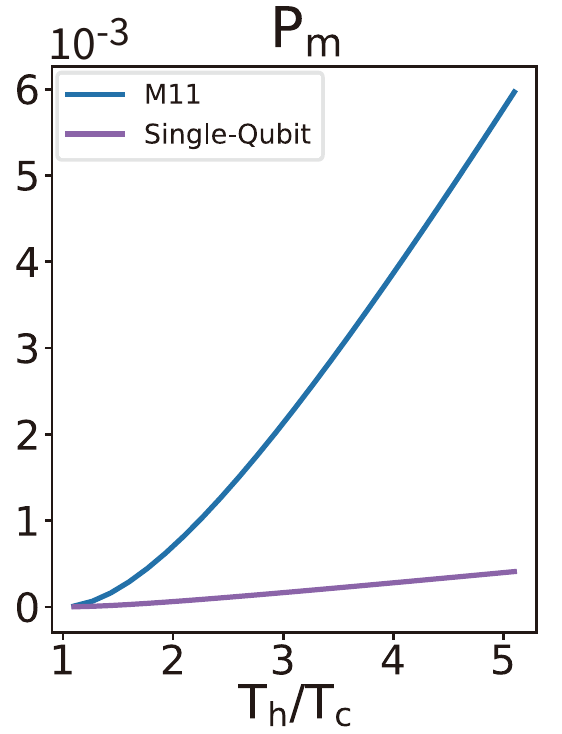}}
	\subcaptionbox[subcaption3]{$(\omega_1^c/\omega_c)_{P_m}$\label{M11:kw}}[0.28\textwidth]{
		\includegraphics[scale=0.77]{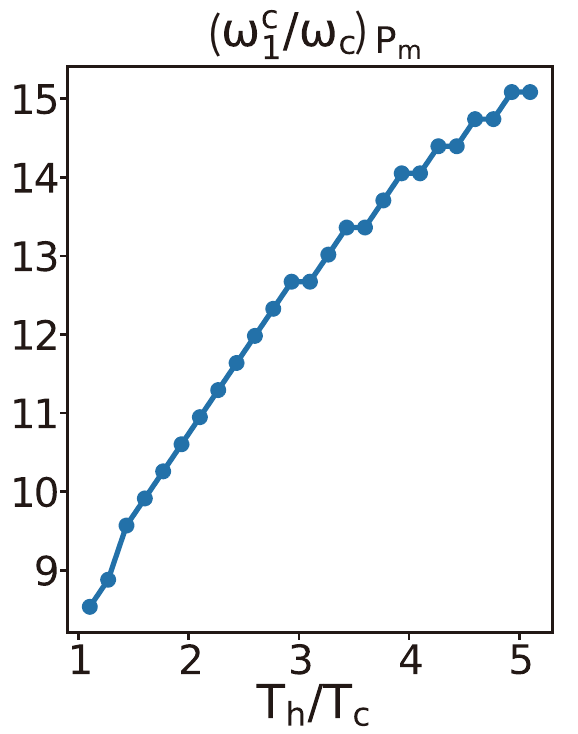}}
	\subcaptionbox[subcaption4]{$\eta_{P_m}$\label{M11:Eff}}[0.33\textwidth]{
		\includegraphics[scale=0.77]{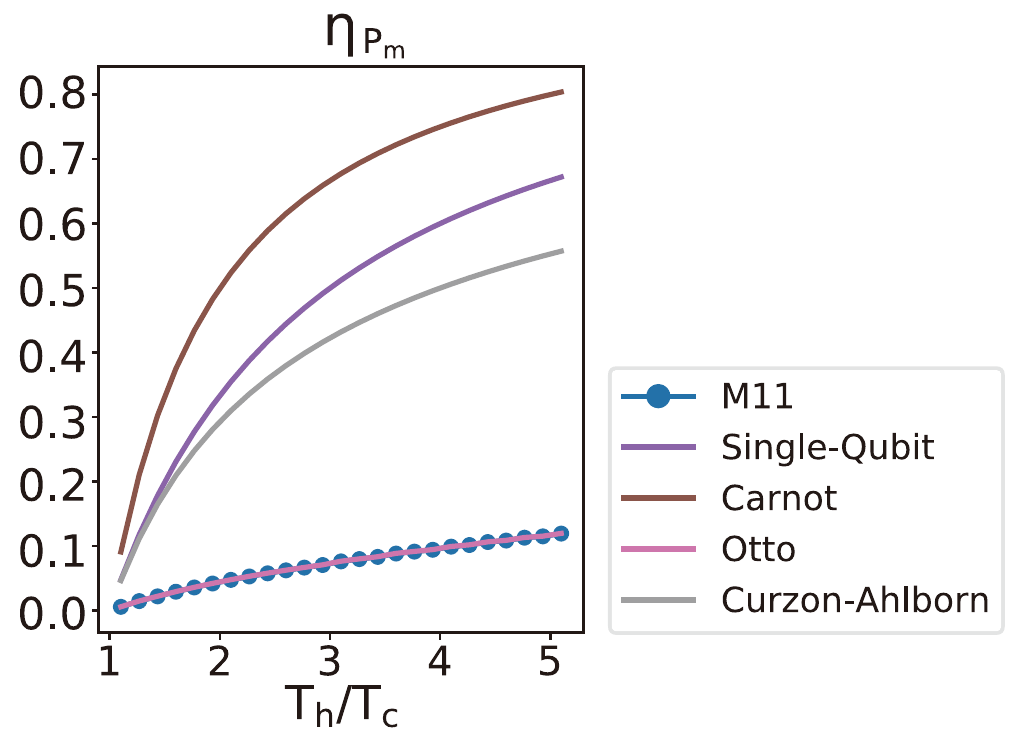}}
	\caption{Dependence of (a)~the maximum power $P_m$, (c)~the energy levels $(\omega_1^c/\omega_c)_{P_m}$ of Q1 and (d)~the efficiency $\eta_{P_m}$ of Model~11 on the heat-bath temperatures $T_h/T_c$ for the fixed coupling strength $g=0.55$. Parameters: energy unit: $\omega_c=1$; temperature of cold bath: $T_c = 5$; transition rates: $\kappa_h = \kappa_c = 0.005$; time durations: $t_h = t_c = 50$.}
	\label{M11_Pm}
\end{figure*}

As shown in Fig.\ref{M11:Pm}, Model~11 achieves much greater powers than the single-qubit one. However, unlike Model~12 and Model~21, which achieve the maximum power for the energy levels of Q1 around two to three times higher than the one that maximizes the power of the single-qubit system, Model~11 achieves the maximum power with much higher energy level $\omega_1^c$ of Q1, over ten times higher than the single-qubit case, as shown in Fig.~\ref{M11:kw}. 
Such a high energy level $\omega_1^c$ of Q1 might not be suitable for many applications. As another point, when the energy levels of Q1 are much higher than the energy level $\omega_2$ of Q2 and the coupling strength $g$ for Model~11, the influence of the coupling to Q2 becomes relatively weak, which is the reason why the maximum power of Model~11 is almost constant independent of the coupling strength. Therefore, although Model~11 achieves the greater power than Model~12, Model~21 and the single-qubit system, we should pay less attention to Model~11. 

As the blue and pink lines that overlap in Fig.~\ref{M11:Eff}, at the maximum power, unlike Model~12 and Model~21, which obtain the system efficiency at the maximum power lower than the Otto efficiency, Model~11 achieves the system efficiency at the maximum power equal to its Otto efficiency, which is similar to the case of the single-qubit engine. In other words, the efficiency of Model~11 at the maximum power is only influenced by the energy levels of Q1, because the energy level $\omega_1^c$ of Q1 that maximizes the power of Model~11 is so high that the influence of the coupling strength becomes trivial. 
Since the energy levels of Q1 at the maximum power are much higher than the case of the single-qubit Otto engine, the efficiency of the Model~11 (blue line) is lower than the single-qubit one (purple line). Different from the single-qubit Otto engine, whose system efficiency (purple line) at the maximum power is greater than the Curzon-Alhborn efficiency (grey line), the system efficiency of Model~11 (blue line) is lower than the Curzon-Alhborn efficiency (grey line).

\begin{figure}
	\centering
	\includegraphics[scale=0.73]{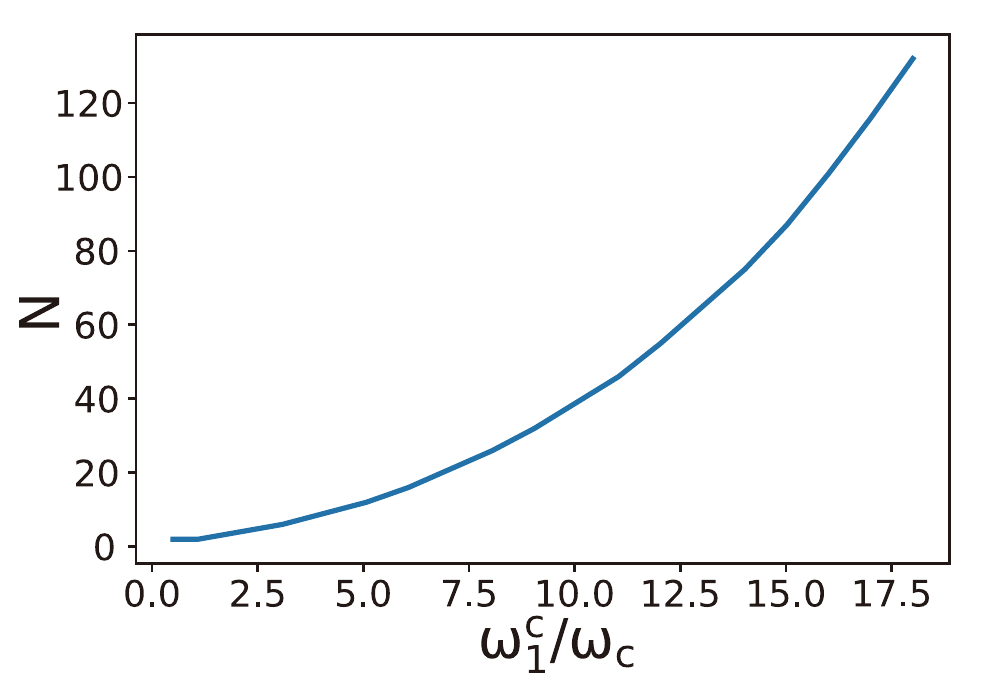}
	\caption{Dependence of the iterations $N$ of Model~11 on the Q1's energy level $\omega_1^c/\omega_c$ for the fixed coupling strength $g=0.4$. Parameters: energy unit: $\omega_c=1$; temperature of each bath: $T_c = 5,T_h=15$; transition rates: $\kappa_h = \kappa_c = 0.005$, time durations: $t_h = t_c = 50$ }
	\label{M11_N}
\end{figure}

Unlike Model~12 and the Model~21, which can always achieve the energy convergence easily, it is difficult for Model~11 to achieve the energy convergence~(\ref{EnergyConvergence}) under some circumstances, so that the iterations $N$ for energy convergence is also a significant factor of Model~11 that we cannot neglect. As shown in Fig.~\ref{M11_N}, if we fix the coupling strength $g$ and the heat-bath temperatures $T_\alpha$, the number of iterations $N$ of the Model~11 to achieve the energy conservation~(\ref{EnergyConvergence}) increases when the energy level $\omega_1^c$ of Q1 becomes big, and becomes significantly large near the maximum-power point, which could be vital in some practical experiments and applications.

\subsection{Model~22}
As Model~22 defined in Sec.~III.~A, the coupled-qubit system contacts with each bath only through Q2 in the ischoric processes, while the work storages interacts with the internal system and produces the work through only Q1.

\begin{figure*}
	\centering
	\subcaptionbox[subcaption1]{Power\label{M22:power}}[0.49\textwidth]{
		\includegraphics[scale=0.75]{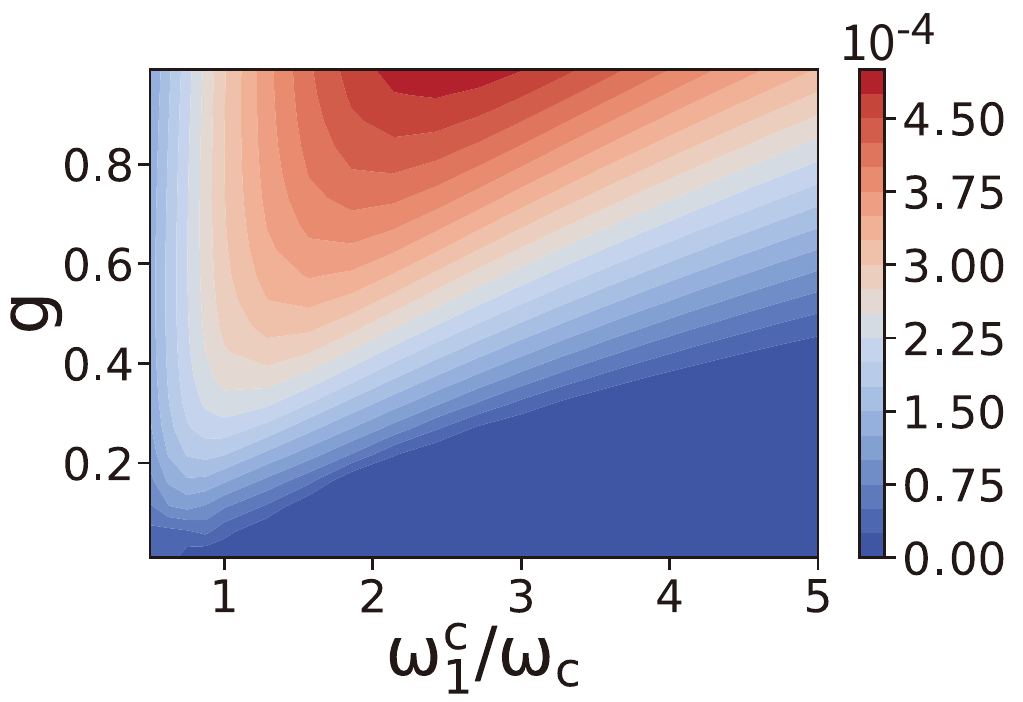}}\hfill
	\subcaptionbox[subcaption3]{Efficiency\label{M22:efficiency}}[0.49\textwidth]{
		\includegraphics[scale=0.75]{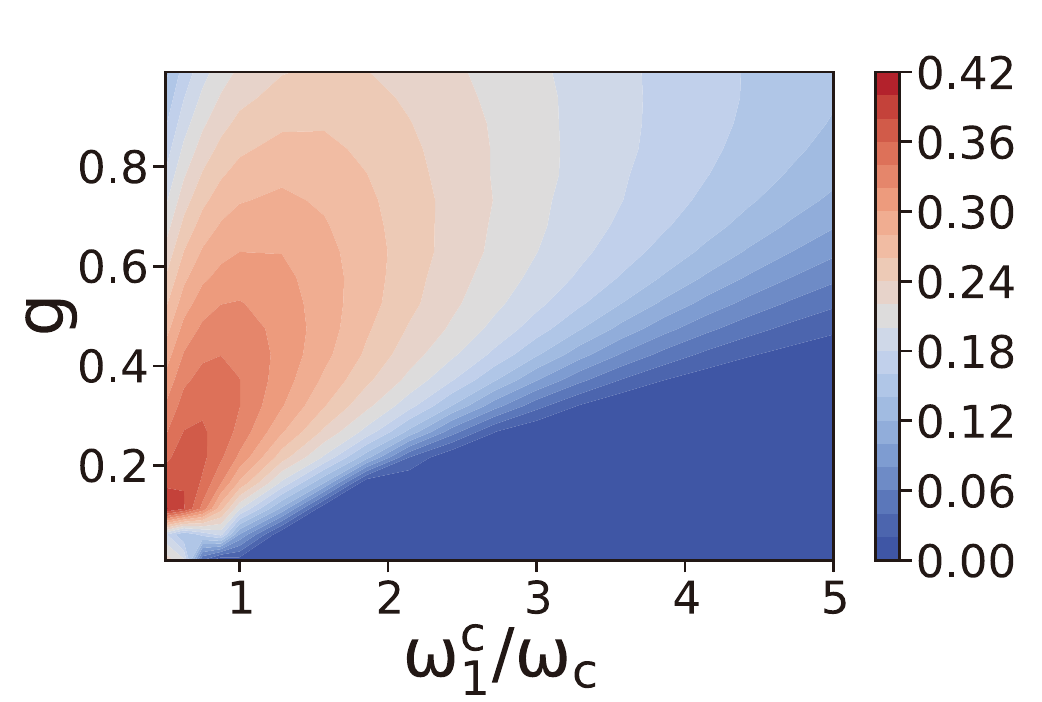}}
	\caption{Dependence of (a)~the power $P$ and (b)~the efficiency $\eta$ of Model~22 on the Q1's energy level $\omega_1^c/\omega_c$ and coupling strength $g$. Parameters: energy unit: $\omega_c=1$; temperature of heat baths: $T_c = 5, T_h=15$; transition rates: $\kappa_h = \kappa_c = 0.005$; time durations: $t_h = t_c = 50$}
	\label{M22_density}
\end{figure*}

\begin{figure*}
	\centering
	\subcaptionbox[subcaption1]{\label{M22:P_kw}}[0.49\textwidth]{
		\includegraphics[scale=0.8]{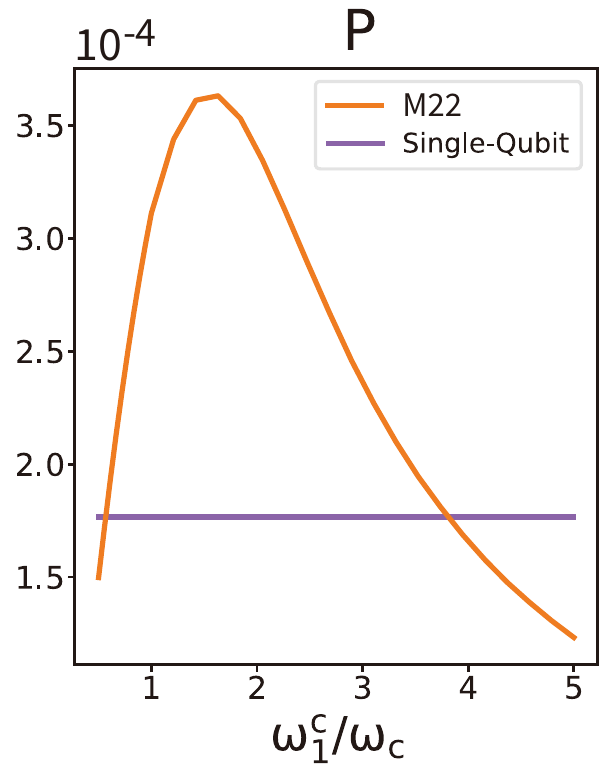}}\hfill
	\subcaptionbox[subcaption2]{\label{M22:Eff_kw}}[0.49\textwidth]{
		\includegraphics[scale=0.8]{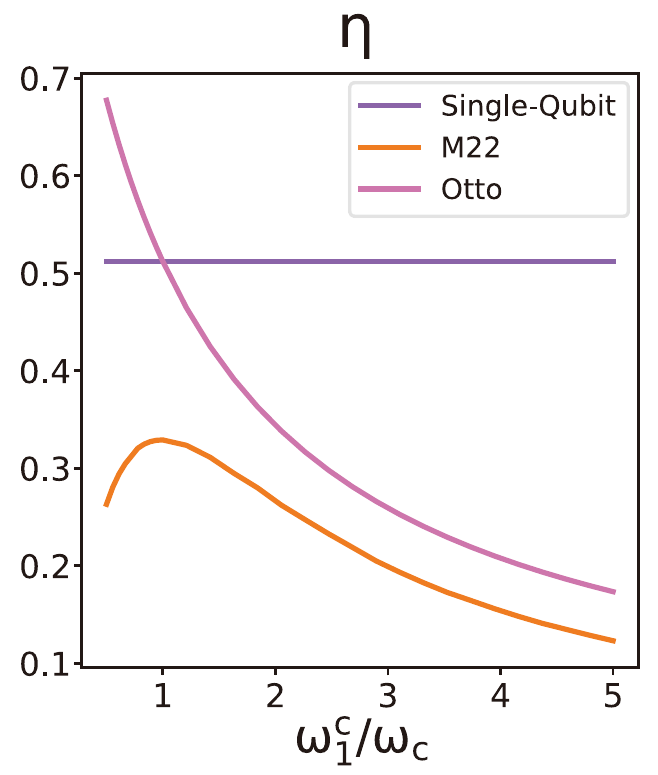}}
	\caption{Dependence of the (a)~power $P$ and (b)~efficiency $\eta$ of Model~22 on the Q1's energy level $\omega_1^c/\omega_c$ for the fixed coupling strength $g=0.55$. Parameters: energy unit: $\omega_c = 1$, temperature of heat baths: $T_c = 5, T_h=15.5$, transition rates: $\kappa_h = \kappa_c = 0.005$; time durations: $t_h = t_c = 50$.}
\end{figure*}

We again search for the maximum power by adjusting the energy level $\omega_1^c/\omega_c$ of Q1 and the coupling strength $g$ under fixed temperatures $T_\alpha$ of the heat baths and the energy-level change $\Delta\omega$ in Eq.~(\ref{MPR3}). The power of Model~22 depends on the coupling strength $g$ and the energy level $\omega_1^c$ of Q1 as shown in Fig.~\ref{M22_density}, which is similar to Model~12 and Model~21; the power of Model~22 increases and approaches to the greatest value when the coupling strength $g$ gets stronger for the fixed energy level $\omega_1^c$ of Q1. 
For a fixed coupling strength, on the other hand, the power of the Model~22 increases first but decreases after a peak when the energy level $\omega_1^c$ of Q1 increases, as shown in Fig.~\ref{M22:P_kw}. We define the peak as the maximum power of Model~22 under specific heat-bath temperatures and coupling strength. As shown in Fig.~\ref{M22:Eff_kw}, the efficiency of Model~22 is lower than its Otto efficiency and the system efficiency of the single-qubit system at the maximum power for the specific heat-bath temperature. In other words, the coupling decreases the efficiency when it improves the power of Model~22, which is similar to Model~12 and Model~21.

\begin{figure*}
	\flushleft
	\subcaptionbox[subcaption1]{$P_m$\label{M22:Pm}}[0.27\textwidth]{
		\includegraphics[scale=0.77]{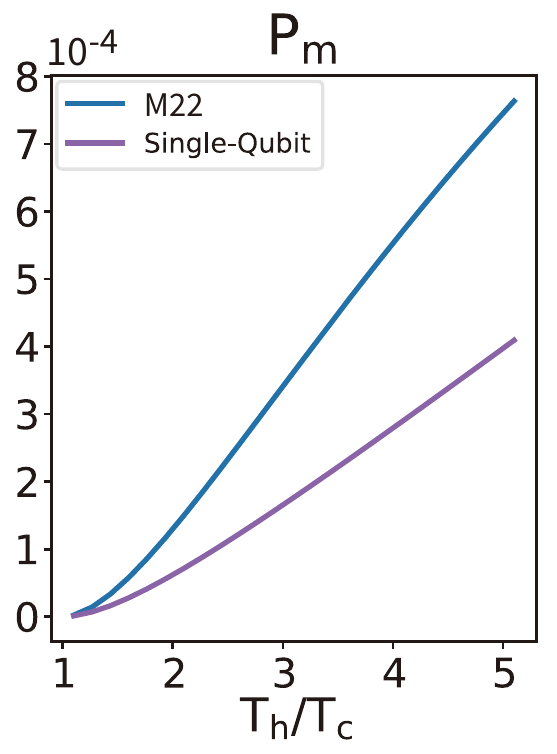}}
	\subcaptionbox[subcaption3]{$(\omega_1^c/\omega_c)_{P_m}$\label{M22:kw}}[0.28\textwidth]{
		\includegraphics[scale=0.79]{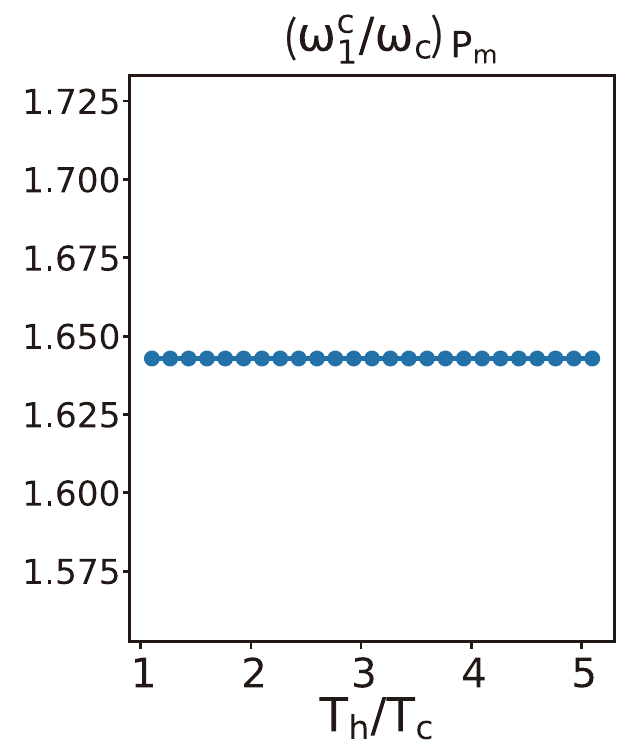}}
	\subcaptionbox[subcaption4]{$\eta_{P_m}$\label{M22:Eff}}[0.33\textwidth]{
		\includegraphics[scale=0.77]{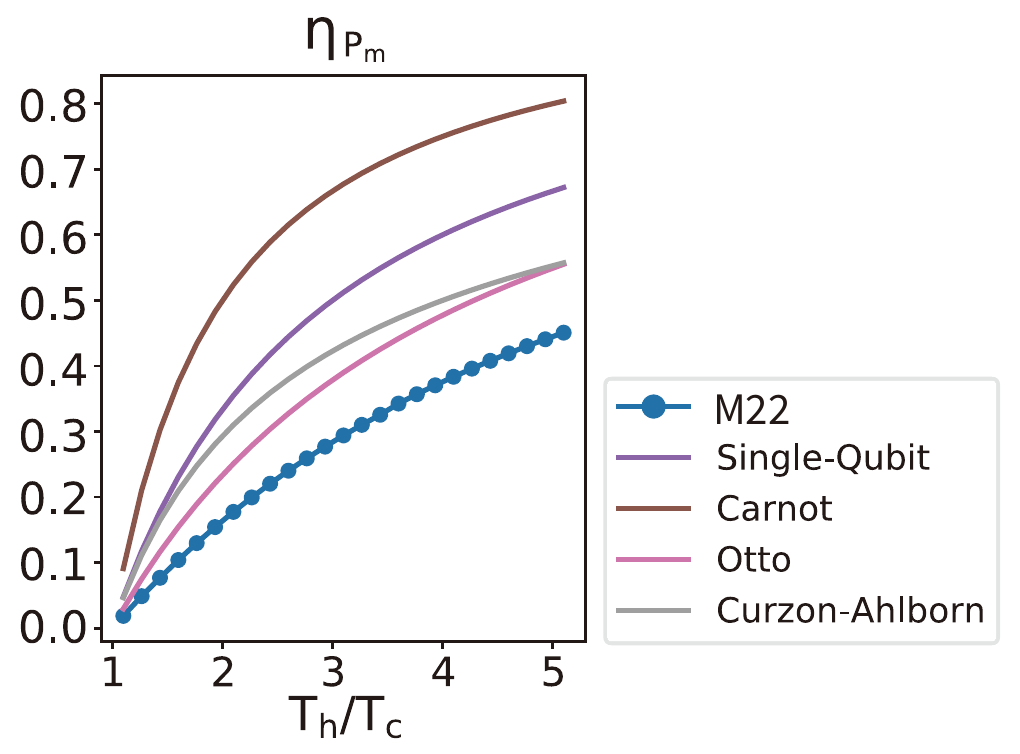}}
	\caption{Dependence of (a)~the maximum power $P_m$, (b)~the energy level $(\omega_1^c/\omega_c)_{P_m}$ of Q1 and (c)~the efficiency $\eta_{P_m}$ of Model~22 on the fixed heat-bath temperatures $T_h/T_c$ for the fixed coupling strength $g=0.55$. Parameters: energy unit: $\omega_c=1$; temperature of cold bath: $T_c = 5$; transition rates: $\kappa_h = \kappa_c = 0.005$; time durations: $t_h = t_c = 50$.}
	\label{M22_Pm}
\end{figure*}

As shown in Fig.~\ref{M22:Pm}, Model~22 yields the maximum power greater than the single-qubit one. Similarly to Model~12, when the cold-bath temperature and the coupling strength are fixed, though the maximum power increases when the temperature of the hot bath increases, the energy level $\omega_1^c$ of Q1 that maximizes the power of Model~22 remains constant for different hot-bath temperature, as shown in Fig.\ref{M22:kw}. 

For the comparison of different efficiencies of Model~22, as the blue line shown in Fig.~\ref{M22:Eff}, similarly to Model~12 and Model~21, the system efficiency is the lowest among the listed efficiencies, due to the influence of the coupling. Because the energy level $\omega_1^c$ of Q1 that maximizes the power of Model~22 is higher than the energy level $\omega_c$ that maximizes the power of the single-qubit system, the system efficiency (blue line) is lower than the single-qubit system's efficiency (purple line) and the Curzon-Ahlborn efficiency (grey line).

\begin{figure}
	\centering
	\includegraphics[scale=0.78]{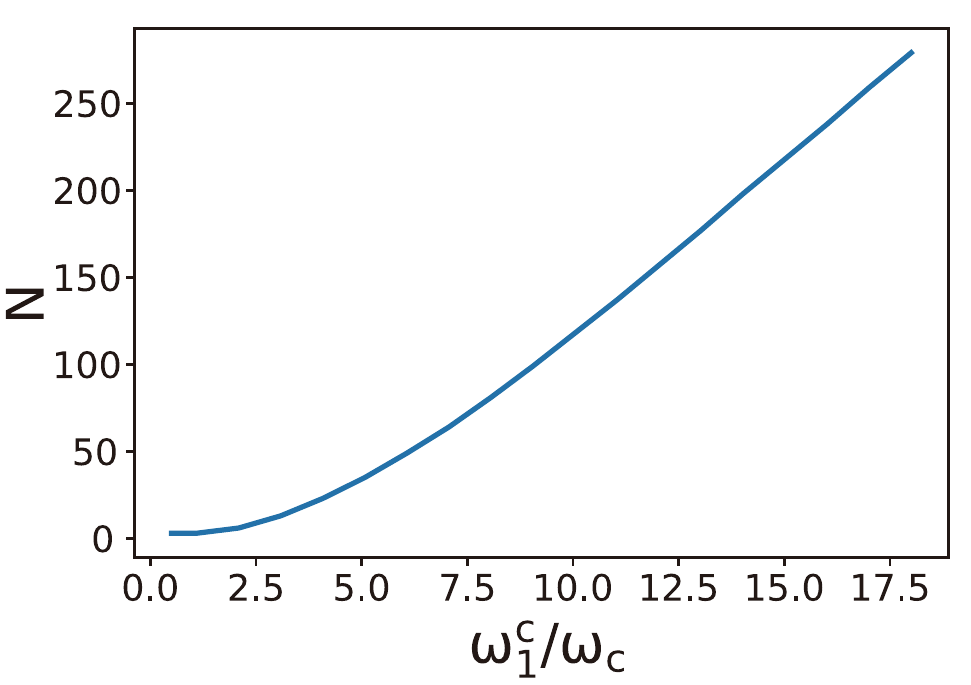}
	\caption{Dependence of the iterations $N$ of Model~22 on the Q1's energy level $\omega_1^c/\omega_c$ for the fixed coupling strength $g=0.4$. Parameters: energy unit: $\omega_c=1$; temperature of each baths: $T_c = 5,T_h=15$; transition rates: $\kappa_h = \kappa_c = 0.005$; time durations: $t_h = t_c = 50$.}
	\label{M22_N}
\end{figure}

Similarly to Model~11, it is difficult for Model~22 to obtain the energy convergence~(\ref{EnergyConvergence}) under some circumstances. The number of iterations $N$ for the energy conservation increases when the energy level $\omega_1^c$ of Q1 becomes higher, as shown in Fig.~\ref{M22_N}, and it becomes difficult for Model~22 to achieve the energy convergence, which is similar to the case of Model~11.

\subsection{Comparision}
Hitherto, from Secs.~V.~A to~D, we analyze the results of the four models of our coupled-qubit Otto engine and verify that the coupled-qubit Otto engine can achieve greater powers than the single-qubit Otto engine in various situations. Let us finally compare these four models of our coupled-qubit Otto engine and the single-qubit system to each other. 
\begin{figure*}
	\flushleft
	\subcaptionbox[subcaption1]{$P_m$\label{Compare:Pm}}[0.26\textwidth]{
		\includegraphics[scale=0.77]{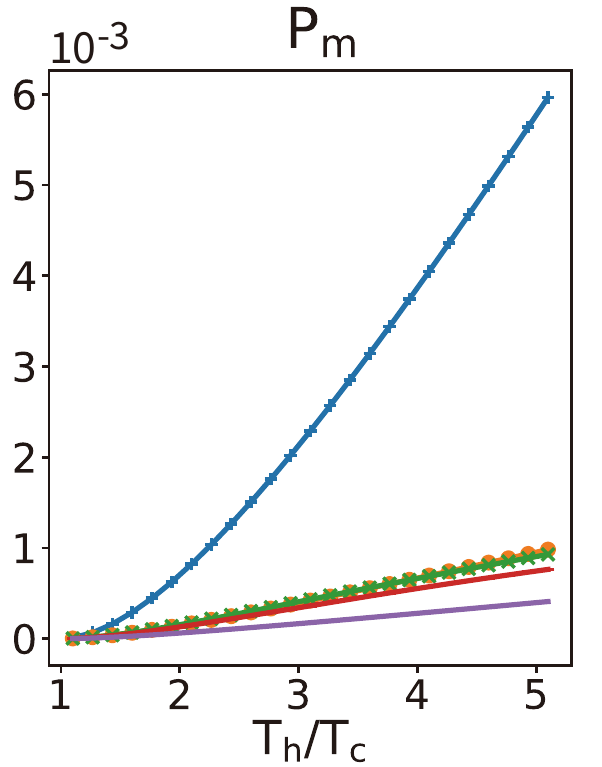}}
	\subcaptionbox[subcaption4]{$\eta_{P_m}$\label{Compare:Eff}}[0.27\textwidth]{
		\includegraphics[scale=0.77]{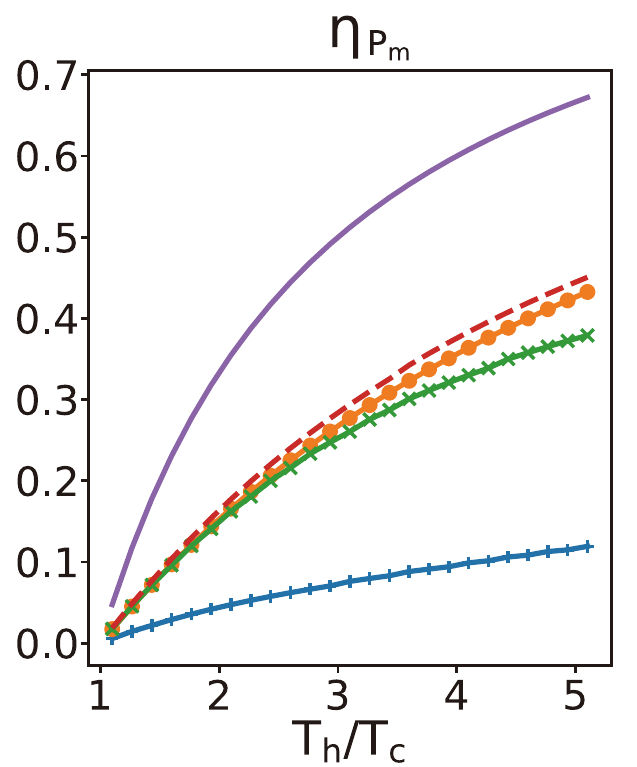}}
	\subcaptionbox[subcaption3]{$(\omega_1^c/\omega_c)_{P_m}$\label{Compare:kw}}[0.31\textwidth]{
		\includegraphics[scale=0.77]{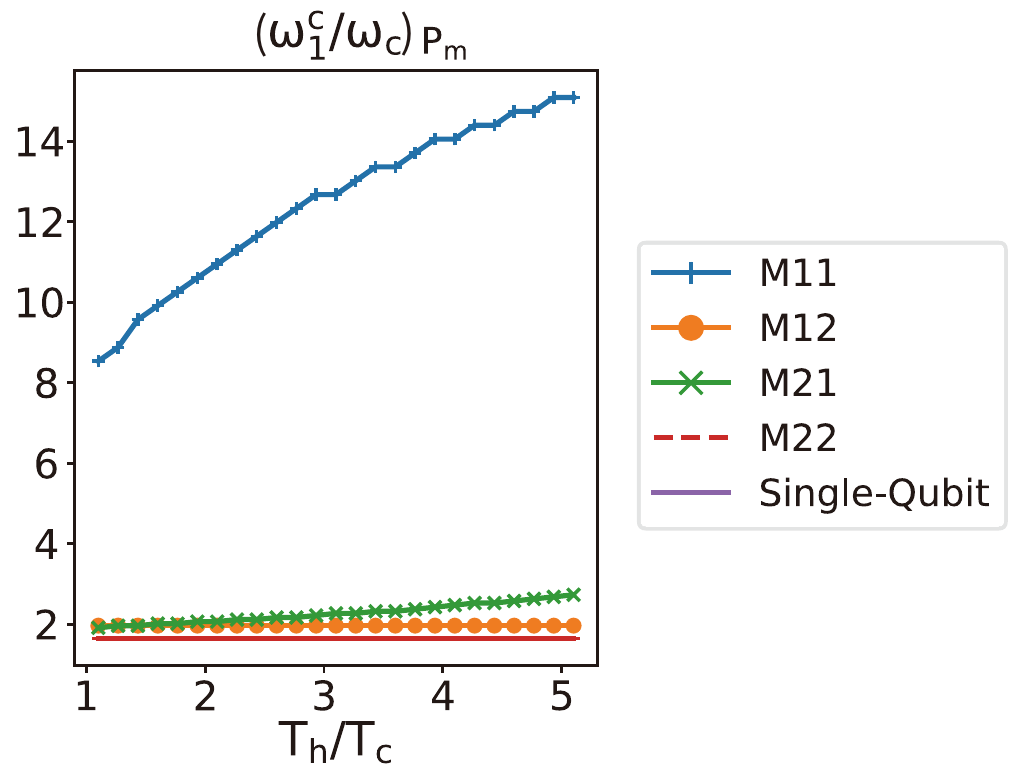}}
	\caption{Dependence of (a)~the maximum power $P_m$, (b)~the efficiency $\eta_{P_m}$ and (c)~the energy levels $\omega_1^c/\omega$ on the heat-bath temperatures $T_h/T_c$ for the fixed coupling strength $g=0.55$. Parameters: energy unit: $\omega_c=1$; temperature of cold bath: $T_c = 5$; transition rates: $\kappa_h = \kappa_c = 0.005$; time durations: $t_h = t_c = 50$.}
	\label{Compare_Pm}
\end{figure*}

For the maximum power obtained by scanning the energy level $\omega_1^c$ of Q1 with fixed coupling strength $g$, as shown in Fig.~\ref{Compare_Pm}, all of the coupled-qubit systems can achieve greater powers than the single-qubit Otto engine. All of our coupled-qubit models break the maximum-power relation~(\ref{MPR1}) of the single-qubit Otto engine and achieve much greater power with a higher energy level $\omega_1^c$ of Q1 than the level $\omega_c$ of the single-qubit system. The power is from the largest to the smallest for Model~11, Model~21, Model~12, Model~22 in this order, and the single-qubit system at last, corresponding to the energy levels of Q1 from the highest to the lowest. 

However, as we mentioned before, although Model~11 achieves the greatest power, its energy level $\omega_1^c$ of Q1 for achieving the maximum power is so high that the influence of the coupling on Model~11 at the point of maximum power is trivial, and such high energy levels might be impractical. Therefore, we mainly focus on Model~12 and Model~21, which achieves the maximum power greater than the one of Model~22 and the single-qubit system. Model~12 and Model~21 achieve the maximum power almost twice of the single-qubit one by the energy level $\omega_1^c$ of Q1 about two to three times higher than the single-qubit case, which is quite practical, and the efficiency at the maximum power is still acceptable comparing to Model~11, whose efficiency at the maximum power is much lower than the other models. 
\begin{figure*}
	\flushleft
	\subcaptionbox[subcaption1]{$P_m$\label{kw1:Pm}}[0.25\textwidth]{
		\includegraphics[scale=0.72]{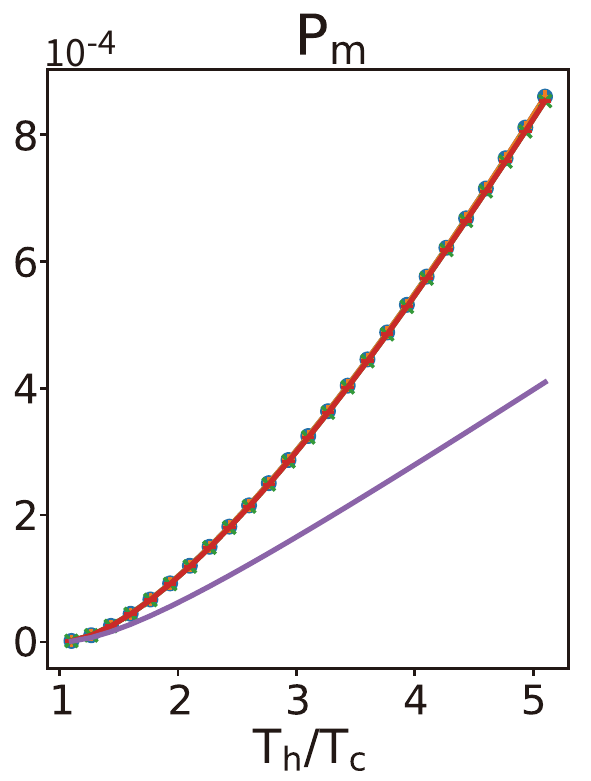}}
	\subcaptionbox[subcaption2]{$g_{P_m}$\label{kw1:g}}[0.27\textwidth]{
		\includegraphics[scale=0.72]{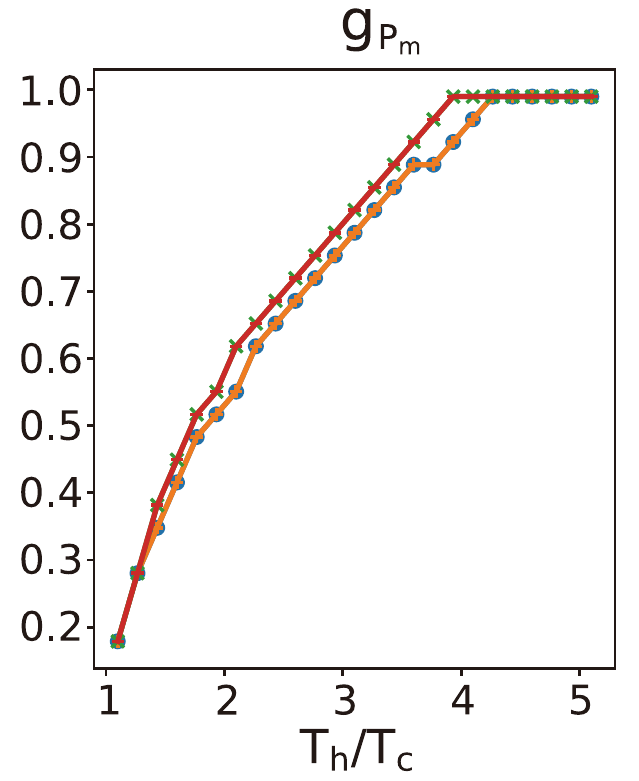}}
	\subcaptionbox[subcaption3]{$\eta_{P_m}$\label{kw1:Eff}}[0.27\textwidth]{
		\includegraphics[scale=0.72]{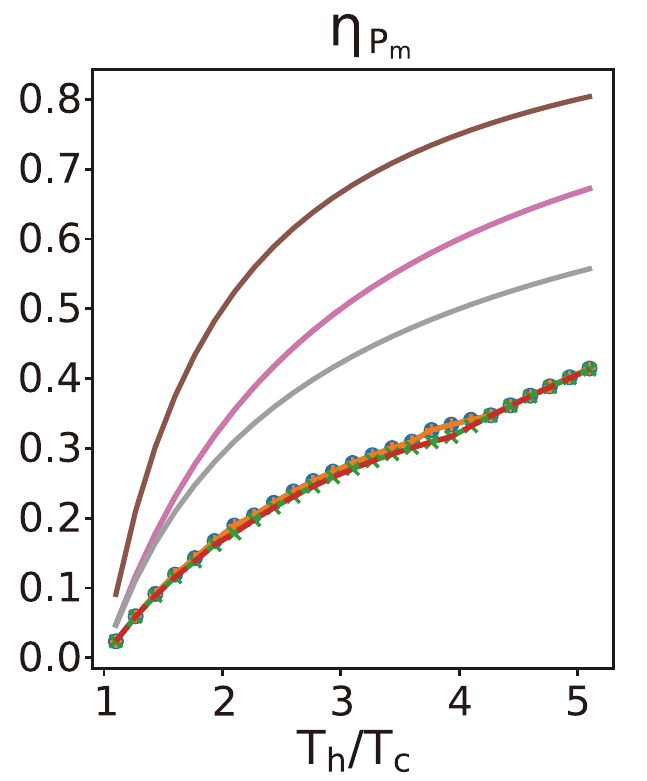}}
	\begin{minipage}[t]{0.13\textwidth}
		\centering
		\includegraphics[scale=0.7]{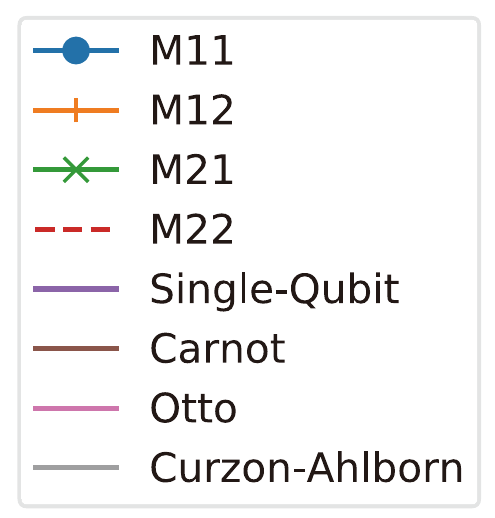}
	\end{minipage}
	\caption{Dependence of (a)~the maximum power $Pm$, (b)~the efficiency $\eta_{P_m}$ and (c)~coupling strength $g_{P_m}$ on the heat-bath temperatures $T_h/T_c$ when the energy levels of Q1 are equal to the single-qubit system. Parameters: energy unit: $\omega_c=1$; temperature of cold bath: $T_c = 5$; transition rates: $\kappa_h = \kappa_c = 0.005$; time durations: $t_h = t_c = 50$.}
	\label{Compare_kw1}
\end{figure*}

For better understanding of the influence of the coupling of the internal system, we search for another maximum power by scanning the coupling strength $g$ and setting other parameters of the coupled-qubit Otto system to the same as the single-qubit one. In other words, we define another type of maximum power depending on the coupling strength $g$ by fixing the energy levels $\omega_1^\alpha$ of Q1 equal to the single-qubit case $\omega_\alpha$, in order to examine the impact of the $XX$-coupling of our system. When the energy levels $\omega_1^\alpha$ of Q1 are equal to the levels $\omega_\alpha$ of the single-qubit system, our coupled-qubit models achieve almost equal maximum power under the similar circumstances, as shown in Fig.~\ref{kw1:Pm}. In the situation, the power of these coupled-qubit systems are greater than the single-qubit case, verifying that the coupling in the coupled-qubit system can improve the power. As shown in Fig.~\ref{kw1:g}, for the energy level $\omega_1^c$ of Q1 which is the same as the single-qubit one, the coupling strength $g$ that maximizes the power of Model~11 (blue-dots) and Model~12 (orange-pluses) are almost equal to each other, and the one that maximizes the power of Model~21 (green-x) and Model~22 (red-dotted line) are almost equal to each other.
For the fixed energy level $\omega_1^c$ of Q1 and the heat-bath temperatures $T_\alpha$, as shown in in Fig.~\ref{kw1:Eff}, depending on different coupling strengths $g$, the system efficiencies of the coupled-qubit systems at the maximum power are lower than the single-qubit one (purple line), and they are lower than their Otto efficiency (pink line) for the specific heat-bath temperature, verifying that the coupling leads the efficiency at the maximum power to decrease while it improves the power.

\section{Conclusions}
In the present paper, we investigate different factors' impacts on the power of several Otto quantum heat engines, comparing the similarity and the difference between the single-qubit and the coupled-qubit systems. For the simulation of these Otto cycles, we utilized the Python toolbox QuTip~\cite{qutip1,qutip2} to calculate the evolution of the systems based on two types of the GKSL master equation~\cite{Hofer_2017, doi:10.1142/S1230161217400108} and the work production processes based on the indirect measurement~\cite{PhysRevResearch.5.023066, PhysRevA.95.032132} with different kinds of work storages. 

For the single-qubit Otto quantum thermal machine, we observe that it can act as three types of thermal machines under diverse heat bath temperatures and system energy gaps. In maximizing the power of the single-qubit engine, we found an almost linear relation~(\ref{MPR1}) between the ratio of the heat baths temperatures $T_h/T_c$ and the ratio of system energy levels $\omega_h/\omega_c$. Utilizing the maximum-power relation, we come up with parameterization of four different models of the coupled-qubit Otto machine with $XX$-coupling. 

We numerically found that the coupled-qubit systems can achieve much greater powers than the single-qubit machine with the same energy-level change. The maximum powers of our coupled-qubit models are also greater than the single-qubit one. The energy levels of the coupled-qubit engines are higher than the single-qubit system when they achieve the maximum power under the specific hath-temperatures and coupling strength, though they are in the same energy-level change as the single-qubit one. 

When the other factors except the coupling strength are the same as the ones the single-qubit system's, a greater power than the maximum power of the single-qubit system is achieved by the coupled-qubit engines, verifying that $XX$-coupling improves the power of the Otto engines. Besides, in all of our coupled-qubit systems, the existence of the coupling to the other qubit in the internal system helps the coupled-qubit Otto engine break the maximum-power relation~(\ref{MPR1}) of the single-qubit system and achieves greater maximum powers with higher energy levels of Q1. Particularly for Model~12 and Model~21, we can achieve much greater powers with practical and reasonable coupling strength and the energy levels of Q1, which could be useful for applications that focus on the power of the quantum Otto engine.
 
Though Model~11 produces the maximum power greater than the other models, the coupling strength influences the value of power trivially at the maximum power. The impact of the coupling on the power and efficiency becomes weak and trivial when the energy levels of Q1 are high, so that the system efficiency is almost equal to the Otto efficiency at the maximum power of Model~11. In addition, the $XX$-coupling of Model~11 makes achieving the energy convergence~(\ref{EnergyConvergence}) difficult, which is critical in practice. 

For the other three models, the influence of the coupling in the power and the efficiency at the maximum power is always significant. We find that their system efficiencies at the maximum powers are lower than their Otto efficiency, unlike the single-qubit system, which yields the system efficiency equal to the Otto efficiency, verifying that the coupling decreases the system efficiency at the maximum power, which is consistent with a trade-off relation between the efficiency and the power~\cite{PhysRevLett.117.190601, PhysRevE.106.024137, PhysRevLett.120.190602, PhysRevE.97.062101, Dechant_2019, Koyuk_2019}.  Due to the higher energy levels of Q1, all the system efficiencies of our coupled-qubit models at the maximum power are lower than the Curzon-Ahlborn efficiency, unlike the single-qubit system, whose efficiency at the maximum power is higher than the Curzon-Ahlborn efficiency.

\section{Acknowledgement}
The present work is partially supported by JSPS KAKENHI Grant Numbers JP19H00658, JP21H01005, and JP22H01140.

\bibliography{reference}
\end{document}